\documentclass[preprint,nofootinbib,aps,superscriptaddress,eqsecnum]{revtex4-1}
\pdfoutput=1
\usepackage[colorlinks=true,citecolor=blue,linkcolor=blue,breaklinks=true]{hyperref}
\usepackage{graphicx}
\usepackage{mathrsfs}
\usepackage{soul}
\usepackage{natbib}
\usepackage{enumerate}
\usepackage{amsmath,amssymb}
\usepackage{slashed}
\usepackage{amssymb,amsmath,subfigure,xcolor}

\newcommand{\lapprox}{%
\mathrel{%
\setbox0=\hbox{$<$}
\raise0.6ex\copy0\kern-\wd0
\lower0.65ex\hbox{$\sim$}
}}
\newcommand{\gapprox}{%
\mathrel{%
\setbox0=\hbox{$>$}
\raise0.6ex\copy0\kern-\wd0
\lower0.65ex\hbox{$\sim$}
}}
\newcommand{\ba}{\begin{array}}
\newcommand{\ea}{\end{array}}
\newcommand{\bd}{\begin{displaymath}}
\newcommand{\ed}{\end{displaymath}}
\newcommand{\beq}{\begin{equation}}
\newcommand{\eeq}{\end{equation}}
\newcommand{\bea}{\begin{eqnarray}}
\newcommand{\eea}{\end{eqnarray}}

\newcommand{\ra}{\rightarrow}
\newcommand{\nn}{\nonumber}

\def\ie{ {\em i.e.,\ }}

\def\a{\alpha}

\def\b{\beta}

\def\m{\mu}
\def\n{\nu}

\def\q2 {q^2}

\def\bt{\begin{table}}
\def\et{\end{table}}

\catcode`@=11 
\def \gsim{\mathrel{\mathpalette\@versim>}}
\def \lsim{\mathrel{\mathpalette\@versim<}}
\def \@versim#1#2{\lower0.4ex\vbox{\baselineskip\z@skip\lineskip\z@skip
     \lineskiplimit\z@\ialign{$\m@th#1\hfil##\hfil$%
     \crcr#2\crcr\sim\crcr}}}
\catcode`@=12 

\begin{document}

\renewcommand*{\thefootnote}{\fnsymbol{footnote}}

\begin{center}

{\large\bf Strong constraints on non-standard neutrino interactions: \\
LHC vs. IceCube}\\[15mm] 
Sujata Pandey\footnote{E-mail:  phd1501151007@iiti.ac.in}, Siddhartha Karmakar\footnote{E-mail: phd1401251010@iiti.ac.in} and Subhendu
Rakshit\footnote{E-mail: rakshit@iiti.ac.in} 
\\[2mm]

{\em Discipline of Physics, Indian Institute of Technology Indore,\\
 Khandwa Road, Simrol, Indore - 453\,552, India}
\\[20mm]
\end{center}

\begin{abstract} 
\vskip 20pt

We find the constraints on various non-standard interactions~(NSI) of neutrinos from monojet+$\slashed{E}_T$ searches at the Large Hadron Collider~(LHC). Also, we show that the measurement of neutrino-nucleon cross-section from the observation of high energy astrophysical neutrino events at IceCube facilitates strong constraints on NSI as well. To this end, we pursue a comparative study of the prospects of LHC and IceCube in detecting NSI, also mentioning the role of low-energy experiments.  
We discuss the case of NSI with a new vector boson $Z'$ and it is found that for some range of $m_{Z'}$ LHC puts more stringent  bound, whereas IceCube supersedes elsewhere. We also pay special attention to the case of  $Z'$ of mass of a few GeVs, pointing out that the IceCube constraints can surpass those from LHC and low-energy experiments. Although, for contact-type  effective interactions with two neutrinos and two partons, constraints from LHC are  superior.

\end{abstract}
 \vskip 1 true cm
 \pacs {}
\maketitle
\section{Introduction}
\label{intro}
The observation of high energy astrophysical neutrinos of extragalactic origin at IceCube~\cite{Aartsen:2015knd,TheIceCube:2016oqi} can provide several useful insights about production mechanism and interactions of such neutrinos. The potential of IceCube in unravelling neutrino decoherence~\cite{Coloma:2018idr, Hooper:2004xr}, existence of sterile neutrinos~\cite{Esmaili:2018qzu,Liao:2016reh, Liao:2018mbg, Moss:2017pur, Denton:2018dqq, Dey:2018yht,  Salvado:2016uqu, Day:2016shw,Esmaili:2013fva}, neutrino interactions with dark matter~(DM)~\cite{Barranco:2010xt, Reynoso:2016hjr, Arguelles:2017atb, deSalas:2016svi, Huang:2018cwo, Kelly:2018tyg, Pandey:2018wvh, Karmakar:2018fno} and cosmic neutrino background~\cite{Ng:2014pca,DiFranzo:2015qea, Araki:2015mya, Mohanty:2018cmq, Chauhan:2018dkd, Shoemaker:2015qul,Cherry:2016jol} have been addressed in the literature.

It is also interesting to ask whether the observation of these high energy neutrinos with $E_{\nu} \gtrsim 20$~TeV provides new information about  neutrino interactions with matter~(partons and electrons). Non-standard  interaction~(NSI) of neutrinos  is a widely studied issue in the literature for several decades. NSIs lead to confusions in extracting the neutrino oscillation parameters from the solar, atmospheric and reactor neutrino data. The non-standard interactions with only one charged lepton, the so-called charged current NSIs, are somewhat constrained from several considerations~\cite{Biggio:2009nt}: CKM unitarity, electroweak precession tests, reactor experiments, {\it etc.} The most stringent bound on $\epsilon_{ll} \bar{\nu_{l}} l \bar{q} q'$ comes from beta decay $\epsilon_{ee} \lsim 4 \times 10^{-4}$~\cite{Bischer:2019ttk}, $\pi^{+}$ semileptonic decay $\epsilon_{\mu \mu} \lsim 4 \times 10^{-3}$, and tau decay $\epsilon_{\tau \tau} \lsim 4.5 \times 10^{-3}$~\cite{Carpentier:2010ue}. Thus, in this paper we do not consider these interactions.

Non-standard interactions with two neutrinos and two partons are relatively less constrained than those of the form $\bar{\nu}_l l \bar{q} q$. Hence, these are the only kind of new interactions we discuss in this paper and refer to as NSI from now on. If these interactions stem from a gauge-invariant operator, interactions with two charged leptons and two partons should also exist, leading to additional constraints from LEP ($e^{+}e^{-} \rightarrow q\bar{q}$)~\cite{Falkowski:2017pss}, muon/tau decay~\cite{Biggio:2009nt}, neutrino-nucleon scattering experiments~\cite{Bergmann:2000gp,Davidson:2003ha, Farzan:2017xzy}, LHC dilepton searches~\cite{Aaboud:2017buh}, {\it etc.} However, it is possible that the new physics~(NP) is such that it leads to NSI with neutrinos, but not  involving their charged counterparts. Later on, we will mention a renormalisable model involving a new vector boson $Z'$ where this situation can be realised. Also, a dimension eight operator, given by $\mathcal{O}_8 =  (\bar{L} H \gamma_{\mu} H^\dagger L)  (\bar{q} \gamma^{\mu} q)$, leads to an interaction of the form $\bar{\nu}\nu \bar{q}q$, but not to its counterpart involving charged leptons $\bar{l} l  \bar{q}q$~\cite{Gavela:2008ra, Davidson:2011kr,Davidson:2003ha}. In this case, the constraints on operators with charged leptons do not apply. Hence, here we do not consider the charged lepton counterpart of interactions with two neutrinos and two partons.

 While neutrino propagates through the earth, NSI in neutrino oscillation can lead to observable signatures at long baseline and reactor experiments, as well as solar neutrino observations and neutrino-scattering experiments: all of these pertain to low-energy constraints on NSI. Even IceCube has placed significant constraints on certain NSI parameters through the observation of atmospheric $\nu_{\mu}$ disappearance at DeepCore~\cite{Aartsen:2014yll,Aartsen:2017xtt}. These constraints are vastly studied in the literature.  However, the observation of high energy  astrophysical neutrinos at IceCube leads to the extraction of neutrino-nucleon deep inelastic scattering cross-sections at unprecedented high  energies~\cite{Aartsen:2017kpd,Bustamante:2017xuy}. The effects of NSI under consideration  lead to signatures similar to  SM neutral current~(NC) neutrino-nucleon scattering at IceCube. Thus, it is possible to constrain the effects of neutrino NSI from the measurement of neutrino-nucleon scattering cross-section at IceCube. 
 
NSI can also show up at the LHC in a final state characterised by missing energy and one or more hard jets. 
Thus the generic search of new physics in channels like monojet+$\slashed{E}_T$ can lead to significant constraints on the NSI parameters. In this paper, we find out the constraints on NSI from the measurement of neutrino-nucleon cross-section at IceCube, as well as LHC monojet+$\slashed{E}_T$ searches and perform a comparative study taking into account the bounds from low-energy neutrino scattering experiments.
Moreover, we consider not only the NSI with $(V - A)$ structure of neutrino currents, but a complete set of interactions up to dim-7  with even more exotic Lorentz structures.

In Section~\ref{exist} we discuss the general structure of NSI interactions and various existing constraints. Specifics of our implementation of the LHC and IceCube constraints are described  in Section~\ref{impli}. Whereas, in Section~\ref{cons} we introduce the non-standard interactions under consideration. Here, we divide the effective interactions into two categories: interactions mediated by a new gauge boson $Z'$ that couples to neutrinos up to dim-5, and contact type interactions up to dim-7. Finally,  we summarise our findings in Section~\ref{concl} and eventually conclude.

\section{NSI and existing constraints}
\label{exist}
Beyond standard model effects in neutrino interactions with other SM fermions can be encoded in higher dimensional effective interactions, the so-called NSIs. 
The constraints on such interactions can be obtained from a wide range of considerations, such as EW precession tests, neutrino oscillation experiments, coherent neutrino-nucleon scattering, colliders, {\it etc.},  and currently at IceCube as well~\cite{Dev:2019anc}. As mentioned earlier, in this paper, we do not consider NSI with a single charged lepton.  We also do not discuss about NSI with two neutrinos and two charged leptons.  
Widely studied `neutral-current' NSI  involving (axial)-vector-like quark and neutrino currents is given as~\cite{Antusch:2008tz,Berezhiani:2001rs},
\bea
\mathcal{L}_\text{NSI} = 2 \sqrt{2} \epsilon^{f C}_{ij} G_{F}  (\bar{\nu}^i \gamma_{\mu} \nu^j)  (\bar{f} \gamma^{\mu} P_C f),
\label{gen}
\eea
where, $\nu$ and $f$ are the SM neutrinos and quarks respectively, and $G_{F}$ is the Fermi coupling constant. Here, $\epsilon^{f C}_{ij}$ is the NSI parameter with $i, j$ as generation indices and $P_{C}$ with $C = L,R$ are the chirality projection matrices. We also use the notation: $\epsilon^{f V}_{ij}= \epsilon^{f L}_{ij} + \epsilon^{f R}_{ij}$. A discussion of key constraints on NSI parameters and a few clarifications related to the present work are appended:

1. {\bf Oscillation experiments}: Interactions in eq.~\eqref{gen}  can also lead to large effects in neutrino oscillation while propagation through matter via the Mikheev-Smirnov-Wolfenstein (MSW) mechanism~\cite{Wolfenstein:1977ue,Mikheev:1986gs}. Such matter effects can show up in the  oscillation data of solar and atmospheric neutrinos, as well as in the long-baseline and reactor experiments. From these neutrino oscillation experiments, the constraints on the parameter $\epsilon$ can vary from $\mathcal{O}(10^{-2})-\mathcal{O}(10^{-1})$ depending on the flavour indices, for example~\cite{Esteban:2018ppq}, $\epsilon_{ee}^{uV} - \epsilon_{\m\m}^{uV} = [-0.020,0.456]$, $\epsilon_{\m\m}^{uV} - \epsilon_{\tau\tau}^{uV} = [-0.005,0.130]$, $\epsilon_{ee}^{dV} - \epsilon_{\m\m}^{dV} = [-0.027,0.474]$, $\epsilon_{\m\m}^{dV} - \epsilon_{\tau\tau}^{dV} = [-0.005,0.095]$, etc.

\vspace{4pt}

2. {\bf Neutrino scattering off electrons and nucleons}: Coherent neutrino-nucleon scattering at COHERENT~\cite{Akimov:2017ade}, coherent neutrino-electron scattering at Borexino~\cite{Borexino:2017fbd} and deep-inelastic neutrino-nucleon scattering at CHARM~\cite{CHARM} can also constrain the NSI parameters. COHERENT imposes the following constraints on NSI parameters at 90$\%$~CL~\cite{Denton:2018xmq}: $\epsilon_{ee}^{qV} = [-0.073,0.023]\oplus[0.16,0.25]$, $\epsilon_{\m\m}^{qV} = [-0.0070,0.033]\oplus[0.15,0.19]$, $|\epsilon_{e\m}^{qV}| \lesssim 0.055$, $|\epsilon_{e\tau}^{qV} |
\lesssim 0.014$ and $|\epsilon_{\m\tau}^{qV}| \lesssim 0.051$, for $q = u,d$. Also, CHARM can provide significant constraints as well~\cite{Altmannshofer:2018xyo}, as example, $\epsilon_{ee}^{uV} = [-0.11,0.27]$, $\epsilon_{\m\m}^{uV} = [-0.03,0.06]$, etc.

\vspace{4pt}

3. {\bf Missing energy signatures at LHC}: At the LHC, the non-standard neutrino interactions can be probed in channels with missing energy in the final state along with one or more jets. From the observation of these channels  at $\sqrt{s}=8$~TeV, the constraints on NSI parameters appearing in eq.~\eqref{gen} read, $\epsilon^{q C}_{ij} \lesssim 0.17$ ~\cite{Friedland:2011za}, whereas the same with $\sqrt{s} = 13$~TeV is given by, $\epsilon^{q C}_{ij} \lsim 0.02$~\cite{Choudhury:2018azm}. These constraints are independent of chirality or neutrino flavour indices of the NSI parameter, as long as the counterpart of eq.~\eqref{gen} involving the charged leptons are not considered. If DM interactions with partons are also present along with NSI, they can also contribute to the process $p p \rightarrow j + \slashed{E}_T$. In such a scenario, the constraints on NSI can be weaker compared to the case when DM-parton interactions are absent. Thus, the new constraints presented in this paper are somewhat conservative, as they indicate the maximum allowed strength of NSI.   

\vspace{4pt}

4. {\bf IceCube}: Observation of atmospheric neutrinos at DeepCore in the energy range $6-56$~GeV suggests that the disappearance of $\nu_{\mu}$ peaks at a neutrino energy, $E_{\nu} \sim 25$~GeV. This gives rise to the following constraint~\cite{Aartsen:2017xtt}, $-0.0067 \lsim \epsilon^{d V}_{\mu \tau} \lsim 0.0081$, which is more stringent than the same from oscillation experiments, $-0.012 \lsim \epsilon^{d V}_{\mu \tau} \lsim  0.009$~\cite{Esteban:2018ppq}. As mentioned earlier, in this paper, we point out that there is another aspect of IceCube observations which can lead to constraints on NSI and has not been addressed in the literature: The measurement of total neutrino-nucleon scattering cross-section~($\sigma_{\nu N}^{tot}$) from the observation of high energy astrophysical neutrinos. Recently, neutrino-nucleon cross-section has been estimated from the shower and track events induced by such high energy neutrinos in refs.~\cite{Bustamante:2017xuy} and ~\cite{Aartsen:2017kpd} respectively.

In brief, the `classical' searches for NSI, such as the neutrino oscillation experiments, constrain the NSI parameters at the  level $\sim \mathcal{O}(10^{-2}) - \mathcal{O}(10^{-1})$. All the experimental constraints discussed above, except LHC and IceCube, deals with  much lower neutrino energies. For instance, at IceCube, the centre-of-mass energy of neutrino-nucleon scattering with $E_{\nu} \sim 400$~TeV comes out to be $\sim 280$~GeV. Also, at  LHC, in the process $p p \rightarrow \nu \bar{\nu} j$, transverse energy of the $\nu \bar{\nu}$-pair typically attains values up to a few hundreds of GeVs. As mentioned earlier, the constraint on Fermi operator-like dim-6 NSI appearing in eq.~\eqref{gen} from LHC is rather significant. Subsequently, measurement of $\sigma_{\nu N}^{tot}$ from observation of high energy neutrinos at IceCube is also expected to place substantial constraints on NSI, due to similar reach in centre-of-mass energy as LHC. Thus, in this paper, we consider the impact of  LHC and IceCube measurements on the NSI up to dim-7, while we also discuss the implications of other lower energy experiments in passing. 
The constraints from low energy neutrino scattering experiments on such NSI from  have been studied in the literature~\cite{Altmannshofer:2018xyo}. Apparently, dim-7 NSI which lead to additional energy enhancement in neutrino-nucleon cross-section compared to the Fermi-type operator in eq.~\eqref{gen}, are even more promising to be detected at IceCube.  We also pay special attention to the case of $Z'$ of mass around a GeV, a well-studied scenario that leads to potentially large NSI effects. As discussed earlier, it is not possible to distinguish the neutrino flavour structure of the NSI parameters at LHC, as the neutrinos of all flavours lead to missing ${E}_T$. In the same way, we extract flavour-independent constraints on NSI from the estimation of neutrino-nucleon cross-section at IceCube.

\section{Implementation of the constraints}
\label{impli}
In the following, we discuss the specifics about the implementation of LHC and IceCube constraints. 

\vskip 10pt

\noindent$\bullet$ {\bf Implementation of monojet+$\slashed{E_{T}}$ constraints from LHC:} 

\noindent Typical search channels of the NSI are characterised by a final state of mono-X (X = jet, $\gamma$) plus missing energy. 
The monojet plus missing transverse energy  signal considered in this paper stems from the process:
\bea
p p \ra \bar{\nu}_{\alpha} \nu_{\b} j, \hskip 15pt  j= q, \bar{q}, g.
\label{pr}
\eea
For the evaluation of cross-section of the above process, we employ \texttt{Madgraph-2.6.1}~\cite{Alwall:2014hca}, which uses the UFO files generated by \texttt{FeynRules-2.3.32}~\cite{Christensen:2008py}. Hadronization of partonic events are performed using \texttt{Pythia-8}~\cite{Sjostrand:2014zea} and hepmc files are created for $\sqrt{s}= 8, 13$~TeV. The hepmc files are passed to \texttt{CheckMATE-2}~\cite{Dercks:2016npn} which checks the compatibility of an interaction against various LHC searches, in our case, the LHC monojet+$\slashed{E}_T$ searches~\cite{ATLAS:2017dnw,Aad:2015zva}. The allowed values of NSI parameters are chosen such that the generated monojet+$\slashed{E}_T$ signal is less than the $95\%$ exclusion limit on the signal.

\vskip 10pt
\noindent$\bullet$ {\bf Implementation of constraints from cascade and track searches at IceCube:} 

\noindent IceCube has observed upgoing as well as downgoing cascade and track events induced by high energy neutrinos~\cite{Aartsen:2013jdh,Aartsen:2013eka}. The upgoing neutrinos travel through the earth to reach IceCube whereas the downgoing neutrinos reach the detector almost uninterrupted. The number of upgoing events is dependent on neutrino flux and neutrino-nucleon cross-section at the detector up to a shadowing factor $S$ encoding the effects of propagation through the earth. The shadowing factor can be evaluated as, $S = \exp[- X(\theta)/ \Lambda(E_{\nu}, \theta)] $, where $X(\theta)$ is the distance travelled through the earth by a neutrino that reaches the IceCube detector from a declination angle $\theta$, where mean free path of neutrinos,  $\Lambda(E_{\nu}, \theta)= m_{N}/ [(\sigma_{NC}+ \sigma_{CC}) \rho(\theta)]$,   with  $m_N$ as mass of the nucleons and $\rho(\theta)$ as average matter density in earth along angle $\theta$.
Thus, for the downgoing neutrinos, the shadowing factor becomes almost unity.  Within an energy interval, neutrinos coming from different directions are distinguished by the shadowing factor, which is also sensitive to neutrino-nucleon cross-section. Thus, it is possible to estimate such cross-section from the observation of high energy astrophysical neutrinos at IceCube.  
Total neutrino-nucleon cross-section has been calculated in this way, taking into account the contained shower events~\cite{Bustamante:2017xuy}. Non-standard interactions as given in eq.~\eqref{gen}, cannot be distinguished from the SM neutral current interaction as both lead to cascade events at IceCube.
Thus, in the presence of an NSI, the total neutrino-nucleon cross-section receives an additional contribution, which in turn leads to a constraint on the NSI parameter.
See Appendix~\ref{implement} for further details.

The neutrino-nucleon interaction at IceCube for neutrino energy greater than $10~$TeV corresponds to the centre-of-mass energy, $\sqrt{s}\gtrsim 140$~GeV. Hence, these neutrinos suffer  deep inelastic scattering~(DIS). The double-differential neutrino-nucleon DIS cross-section of such interaction is given by~\cite{Gandhi:1998ri}:
\bea
\frac{d^{2} \sigma_{\nu N}}{dx dy} =   \frac{|\mathcal{M}_{\nu q}|^{2}}{16 \pi x s} \Big(f_{q}(x, Q^2)+(1-y)^{2} f_{\bar{q}}(x, Q^2)\Big).
\label{nutnu}
\eea
Here, $x$, $y$ are Bjorken scaling parameters, while $Q$ is the momentum transferred to the nucleon. $f_{q,\bar{q}}(x, Q^2)$  are certain combinations of parton distribution function~(PDF) of the quarks and antiquarks: 
\begin{center}
$f_{q}=(f_{u}+f_{c}+f_{t})~L_{u}^{2}+(f_{\bar{u}}+f_{\bar{c}}+f_{\bar{t}})~R_{u}^{2}+(f_{d}+f_{s}+f_{b})~L_{d}^{2}+(f_{\bar{d}}+f_{\bar{s}}+f_{\bar{b}})~R_{d}^{2}$,\\ 
$f_{\bar{q}}=(f_{u}+f_{c}+f_{t})~R_{u}^{2}+(f_{\bar{u}}+f_{\bar{c}}+f_{\bar{t}}) L_{u}^{2}+(f_{d}+f_{s}+f_{b})~R_{d}^{2}+(f_{\bar{d}}+f_{\bar{s}}+f_{\bar{b}})~L_{d}^{2}$,\\
\end{center}
with,
\bea
L_{u}&=&1/2-2/3 ~\sin^{2} \theta_{W}, \hspace{15pt}
 L_{d}=-1/2+2/3~ \sin^{2} \theta_{W}, \nn\\
R_{u}&=&-2/3~ \sin^{2}\theta_{W}, \hspace{37pt} R_{d}=1/3~ \sin^{2}\theta_{W}.
\eea
Here, $f_{i(\bar{i})}$ are the individual PDFs for the quarks and antiquarks, with $ i = u, d, c, s, t, b$.
 In eq.~\eqref{nutnu}, $|\mathcal{M}_{\nu q}|^{2}$ is the square of amplitude for a given neutrino-parton interaction. For the standard model, the differential cross-section is given as:
\bea
\frac{d^{2} \sigma_{\nu N}}{dx dy} =  \frac{2 G_{F}^{2} m_{N} E_{\nu} x}{ \pi} \frac{m_{V}^{4}}{(m_{V}^{2}+Q^{2})^{2}} \Big(f_{q}(x, Q^2)+(1-y)^{2}f_{\bar{q}}(x, Q^2)\Big),
\label{nutnuSM}
\eea
 where $m_{V} = m_{Z}  (m_W)$ for SM NC~(CC) interactions respectively. In order to calculate $\sigma_{\nu N}$ for neutrino energy $E_{\nu}$, the  PDFs are required to be known in the $x$-range $\{x_{min},1\}$ with, $x_{min} \sim Q^2/(2 m_{N} E_{\nu}) \sim m_{V}^2/ (2 m_{N} E_{\nu})$.
Thus, evaluation of neutrino-nucleon cross-section for neutrino energies from TeV to PeV requires  knowledge of PDFs evaluated at $x \gsim 10^{-4}$. 
These are known from $e p$ collisions at HERA~\cite{Aaron:2009aa, Abramowicz:2015mha} for $x \gsim 2 \times 10^{-5}$. 
Also, LHCb significantly reduces the uncertainties in PDF for such small values of $x$~\cite{Zenaiev:2015rfa, Gauld:2015yia, Cacciari:2015fta}. Hence, the uncertainties in the  neutrino-nucleon cross-sections stemming from QCD effects are rather small. Therefore for high energy astrophysical neutrinos, any significant difference between predicted SM cross-section and the cross-section measured from the observation of IceCube events can be attributed to non-standard interactions. 
We use the CT10 parton distribution functions~\cite{Lai:2010vv} in this work. 

\section{Constraints on NSI interactions from LHC and IceCube}
\label{cons}
 
The NSI interactions can be generated in various extensions of the SM. A complete set of higher dimensional effective  interactions of neutrinos with partons up to dim-7 have been constructed in the literature~\cite{Altmannshofer:2018xyo}. In this section, we consider these effective interactions up to dim-7 which can give rise to neutrino-nucleon scattering at IceCube. Here, we investigate and compare the constraints on the NSI parameters from the neutrino-nucleon cross-section measurement facilitated by IceCube and monojet+$\slashed{E}_T$ search at LHC.
The following discussion is separated in two parts: ($i$) the case of a $Z'$ of mass $\sim \mathcal{O}$(GeV) with renormalisable and effective couplings to neutrinos and quarks, leading to non-standard effects in neutrino-nucleon scattering, and ($ii$), the case of contact type NSI interactions. For the second part, we consider effective operators leading to NSI up to dim-7.

\vspace{10pt}

\subsection{$Z'$ with renormalisable and effective coupling to neutrinos}
\label{Z'eft}
\noindent The NSI generated from a new vector boson coupling to both neutrino and quark currents is particularly important as it can lead to sizable NSI parameters which can be tested at neutrino oscillation and scattering experiments~\cite{Miranda:2015dra, Ohlsson:2012kf, Farzan:2015doa}. 
Such a new vector boson $Z'$ can be realised as the gauge boson corresponding to a $U(1)$ symmetry, pertaining to various chiral anomaly-cancelling combinations of baryon and lepton numbers, for example, $B-L$. The coupling of $Z'$ can even violate lepton flavour universality when realised as the gauge boson corresponding to, for example, $U(1)_{L_{\m} - L_{\tau}}$~\cite{Altmannshofer:2019zhy}, {\it etc.} But, as mentioned earlier, all the constraints on NSI derived in this paper are flavour-independent. 

\vspace{10pt}

\noindent {\bf Constraints on a light $Z'$}: Here we briefly discuss the key constraints on a $Z'$ of mass in the range MeV to GeV and tree-level coupling with neutrinos and quarks, from low-energy experiments and cosmological considerations. If a $Z'$ in the aforementioned mass range couples to charged leptons at the tree-level, several other constraints ensue, which do not apply in our context.  

1.  $Z'$ with tree-level coupling to neutrinos may keep the neutrinos in equilibrium with photons, and thus electrons, even after the thermal decoupling of neutrinos from the rest of the SM particles, which occurs at temperature $T_\text{dec} \sim 2$~MeV in standard cosmology.
This might contradict the measurement of effective numbers of neutrinos~($N_{\rm{eff}}$) and the ratio $Y_\text{He}/Y_\text{H}$ at the BBN epoch, which makes $Z'$ with masses  $m_{Z'} \lesssim 5$~MeV unfavourable~\cite{Huang:2017egl}.

2. Coupling of neutrinos to $Z'$ can lead to non-standard effects in supernova cooling and could leave potential signatures in the observed spectrum of supernova neutrinos, which  constrains  the $Z'$ coupling to be  as small as $g_{\nu} \sim 10^{-10}$ depending upon $m_{Z'}$. This constraint is not applicable for $m_{Z'} \gtrsim 30$~MeV~\cite{Dent:2012mx,Harnik:2012ni}.

3. In the presence of tree-level couplings to quarks, $Z'$ can have kinetic mixing with photons and consequently, several constraints from meson decay apply. For $m_{Z'} = 100 - 200$~MeV, the measurement of the branching ratio for $K_{L}^0 \rightarrow \pi^0 Z'$ leads to the bound $g_q \lesssim 10^{-8}$~\cite{Nelson:1989fx}.  For even lower masses of $Z'$ up to a MeV, this constraint is even more stringent. In the range, $m_{Z'} = 200 - 600$~MeV, measurements related to decays of $\eta$, $\eta^{\prime}$, $\phi$ put bounds on the $Z'$ coupling, with the measurement of $\eta \rightarrow \pi^{0} \gamma \gamma$ providing the most stringent limit of $g_q \lesssim 10^{-5} - 0.01$ depending upon $m_{Z'}$~\cite{Tulin:2014tya}. Measurements of branching ratios of $\eta' \rightarrow \pi^{0} \pi^{+} \pi^{-} \gamma$, $\Psi \rightarrow K^{+}K^{-}$ and $\Upsilon \rightarrow $ hadrons provide comparably weaker constraints, $g_q \lesssim 0.01 - 0.1$ for very narrow ranges of $Z'$ mass at $m_{Z'} = 0.8, 5.5 $ and 9.8 GeV respectively, which correspond to masses of the decaying mesons.

4. For $m_{Z'} \lsim 10$~GeV, BABAR puts a constraint on the electrons-$Z'$ coupling from the measurement of $e^{+} e^{-} \ra \gamma Z'$~\cite{Aubert:2008as,Essig:2013vha} which reads  $g_{e} \lsim 3.3 \times 10^{-2}$. Though in our scenario $Z'$ does not couple to electrons at the tree-level and $e^{+} e^{-} \ra \gamma Z'$ only occurs at one-loop level. Thus, in our case this constraint applies up to a loop factor, significantly downsizing its relevance. This will be addressed in details later.

5. For $Z'$ with tree-level couplings to neutrinos and electrons, Borexino provides significant constraints for $m_{Z'}$ up to a few GeVs. For $m_{Z'} \sim 1$~GeV, this constraint is given by $g_{e,\mu} \lesssim \mathcal{O}(10^{-2})$~\cite{Harnik:2012ni}. For smaller values of $m_{Z'}$, such a constraint can be even more stringent, as an example, $g_{e,\mu} \lesssim \mathcal{O}(10^{-5})$ for $m_{Z'} \sim 1$~MeV~\cite{Harnik:2012ni}. But similar to the last point, for our case, this constraint is  not  that relevant as the neutrino-electron scattering suffers a loop suppression.

In light of the above discussions, broadly the constraints on tree-level couplings of $Z'$ to neutrinos and quarks for $m_{Z'} \lesssim 1$~GeV are quite stringent, owing to the decays of various mesons, cosmological/astrophysical observations, etc. On the other hand, as it will be discussed in details later, for $m_{Z'} \gtrsim 100$~GeV, constraints from LHC on such a  $Z'$ are significant as well, $\epsilon \equiv g_{q} g_{\nu} (v^2/2 m_{Z'}^2)  \sim 0.01$. 
Though, for $Z'$ mass of a few GeVs, $Z'$ couplings remain essentially unconstrained from both the low-energy experiments and LHC, keeping aside the  constraints from $\Psi$ and $\Upsilon$ decay which affect  only small $m_{Z'}$ ranges around the corresponding meson masses.
This situation arises because in order to enable detection of generic new physics signatures, the minimum value of missing $E_T$ at LHC is  considered to be $\gtrsim 100$~GeV, whereas the highest energy reach of the relevant low-energy experiments is up to a GeV.

In the following, we study the cases of a $Z'$ of mass $\sim \mathcal{O}$(GeV), with renormalisable and effective coupling to neutrinos up to dim-5.   

\begin{enumerate}
\item 
Here we consider the renormalisable $Z'$ interaction terms leading to a tree-level neutrino-quark scattering, 
\bea
\mathcal{L} \supset  g_{\nu}(\bar{\nu}  \gamma_{\mu} P_L \n)Z'^{\mu} + g_{q} (\bar{q} \gamma^{\mu} q) Z'_{\mu}.
\label{op1}
\eea
 As it was mentioned earlier, we do not consider the couplings of $Z'$ with charged leptons at the tree-level. Such a scenario can be realised in renormalisable models~\cite{Pandey:2018wvh}, where the $Z'$ is realised as the gauge boson corresponding to an additional $U(1)$ symmetry, under which SM quarks, neutrinos, and the new fermions~($F$) required for cancelling chiral anomalies, transform non-trivially.
Thus, as it can be followed from eq.~\eqref{op1}, the quark couplings with $Z'$ lead to a kinetic mixing of $Z'$ with photon, $\mathcal{L}_{mix} = \epsilon_{loop} F_{\m\n} Z^{\prime \m\n}$, with the following mixing factor, 
\bea
\epsilon_{loop}  \sim \frac{8}{9} \frac{e g_q}{(4 \pi)^2} \ln\Big[\frac{(m_{u} m_{c} m_{t})^2} {(m_{d} m_{s} m_{b}) m_F^3}\Big] = 1.3 \times 10^{-2} ~g_{q} \ln\Big[\Big(\frac{100~\text{GeV}}{m_F}\Big)^3\Big].
\label{loopfac}
\eea
Here, $m_{q}$ is the mass of the quarks, $q = u, c, t, d, s, b$.
Masses of the new fermions can be constrained from several LEP searches as, $m_F \gtrsim 100$~GeV~\cite{Achard:2001qw}. Due to the loop-induced mixing of $Z'$ and $\gamma$, in our scenario, the amplitude of neutrino-electron scattering in Borexino is suppressed by  $\epsilon_{loop}$. Thus, in our case, the constraint from Borexino turns out to be,   $g_q g_{\nu} \lsim 0.25$ for  $m_{Z'}=5~\text{GeV}$, and is relaxed compared to the case of a $Z'$ with tree-level coupling to electrons.  A similar discussion holds for a gauged $L_{\mu}- L_{\tau}$ model~\cite{Araki:2015mya,Abdullah:2018ykz}.

The constraint from LHC monojet+$\slashed{E_{T}}$ searches~\cite{Aad:2015zva} at $\sqrt{s} = 8$~TeV on the interaction in eq.~\eqref{op1} comes out to be, $g_{q} g_{\nu}\lsim 9.9 \times 10^{-3}$ for $m_{Z'} = 5$~GeV.   Whereas,  for the same $Z'$ mass, the constraint from  LHC search~\cite{ATLAS:2017dnw} in the same channel at  $\sqrt{s} = 13$~TeV is weaker, $g_{q} g_{\nu}\lsim 1.7 \times 10^{-2}$. 
This occurs because of a larger background and cuts at larger values of $\slashed{E}_T$ at $\sqrt{s} = 13$~TeV compared to $\sqrt{s} = 8$~TeV  for the process under consideration, leading to a smaller signal-to-background ratio  when the 13~TeV data is adopted. 
In the process $ p p \rightarrow \nu \bar{\nu} j$ at LHC,  the subprocess $q g \rightarrow \nu \bar{\nu} j$ dominates over the $q \bar{q}$ initiated process, due to a large gluon flux. 
Anyway, this implies, for $m_{Z'} \sim 5$~GeV and interactions as in eq.~\eqref{op1}, the LHC constraint at $\sqrt{s} = 8$~TeV is more significant compared to the Borexino bound. However, as it can be seen from fig.\,\ref{fig:simpleZp}, the IceCube observation of the cascade events give a slightly better bound than the LHC monojet+$\slashed{E_{T}}$ searches, $g_{q} g_{\nu}\lsim 1.65 \times 10^{-3}$,\ie $\epsilon \equiv g_q g_{\nu} (v^2/2 m_{Z'}^2) \lsim 2.0$ for $m_{Z'} = 5$~GeV. The maximum allowed values of $\epsilon$ can be read off fig.\,\ref{fig:MZP}\,(a) for different values of $m_{Z'}$.
As it can be seen from fig.~\ref{fig:MZP}~(a), for the interaction in eq.~\eqref{op1}, except the range $m_{Z'} \sim 35 - 500$~GeV, IceCube provides a better constraint than LHC. This happens due to LHC's rather good acceptance in the  channel $ p p \rightarrow j + \slashed{E}_T$  for the aforementioned $Z'$ mass range with renormalisable interactions~\cite{Friedland:2011za}. In this case, the dependence of the LHC constraint on $m_{Z'}$ is similar to that previously found in the literature~\cite{Franzosi:2015wha}.

\begin{figure}[h!]
 \begin{center}
 \includegraphics[width=3.7in,height=2.5in, angle=0]{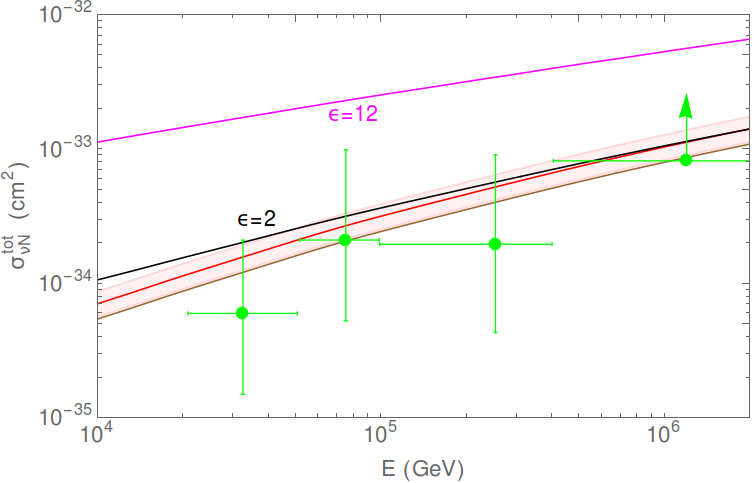}
 \caption{Constraints on NSI described by eq.~\eqref{op1} for $m_{Z'} = 5$~GeV. The brown line represents the total SM  neutrino-nucleon cross-section~\cite{Gandhi:1998ri}. The red line and the light red band denote the central value and 1\,$\sigma$ allowed range of $\sigma_{\nu N}^{tot}$ from IceCube observation of track events respectively~\cite{Aartsen:2017kpd}.
Similarly, the green points and related error bars in the $y$-direction stand for the central values and 1\,$\sigma$ allowed ranges in $\sigma^{tot}_{\nu N}$ measured from the IceCube observation of shower events at different energy bins respectively~\cite{Bustamante:2017xuy}. In presence of a $Z'$ with mass $m_{Z'} = 5$~GeV and interactions as in eq.~\eqref{op1}, ($i$)~the magenta line depicts the value of  $\sigma_{\nu N}^{tot}$ with the  NSI parameter $\epsilon$ set at its maximum allowed value from LHC, $\epsilon = 12$, ($ii$)~the black line represents the value of $\sigma^{tot}_{\nu N}$ with $\epsilon$ set at its maximum allowed value from IceCube, $\epsilon = 2$. }

 \label{fig:simpleZp}
 \end{center}
 \end{figure}

\begin{figure}[h!]
 \begin{center}
\subfigure[]{
 \includegraphics[width=2.6in,height=2.2in, angle=0]{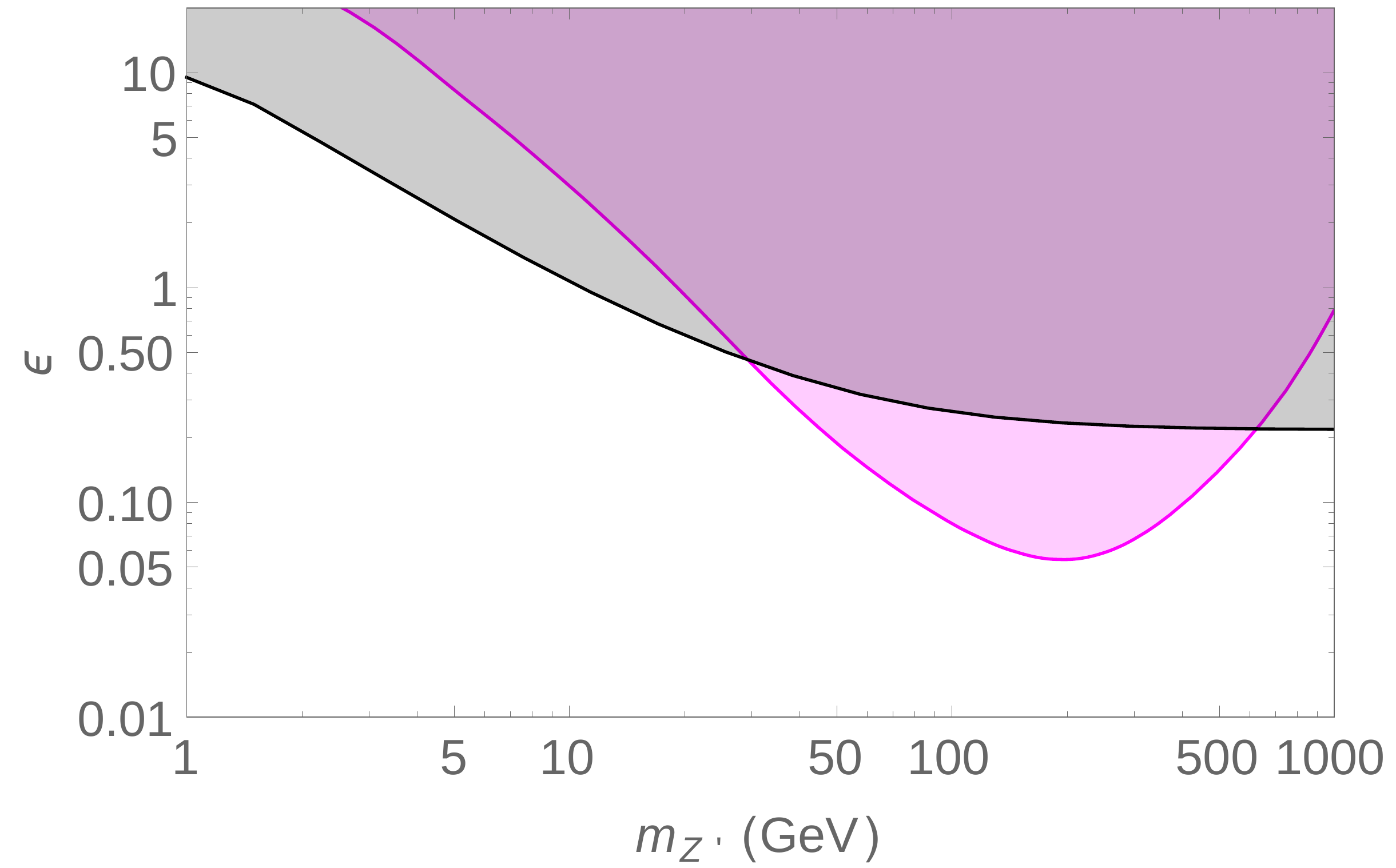}}
 \hskip 15pt
 \subfigure[]{
 \includegraphics[width=2.6in,height=2.2in, angle=0]{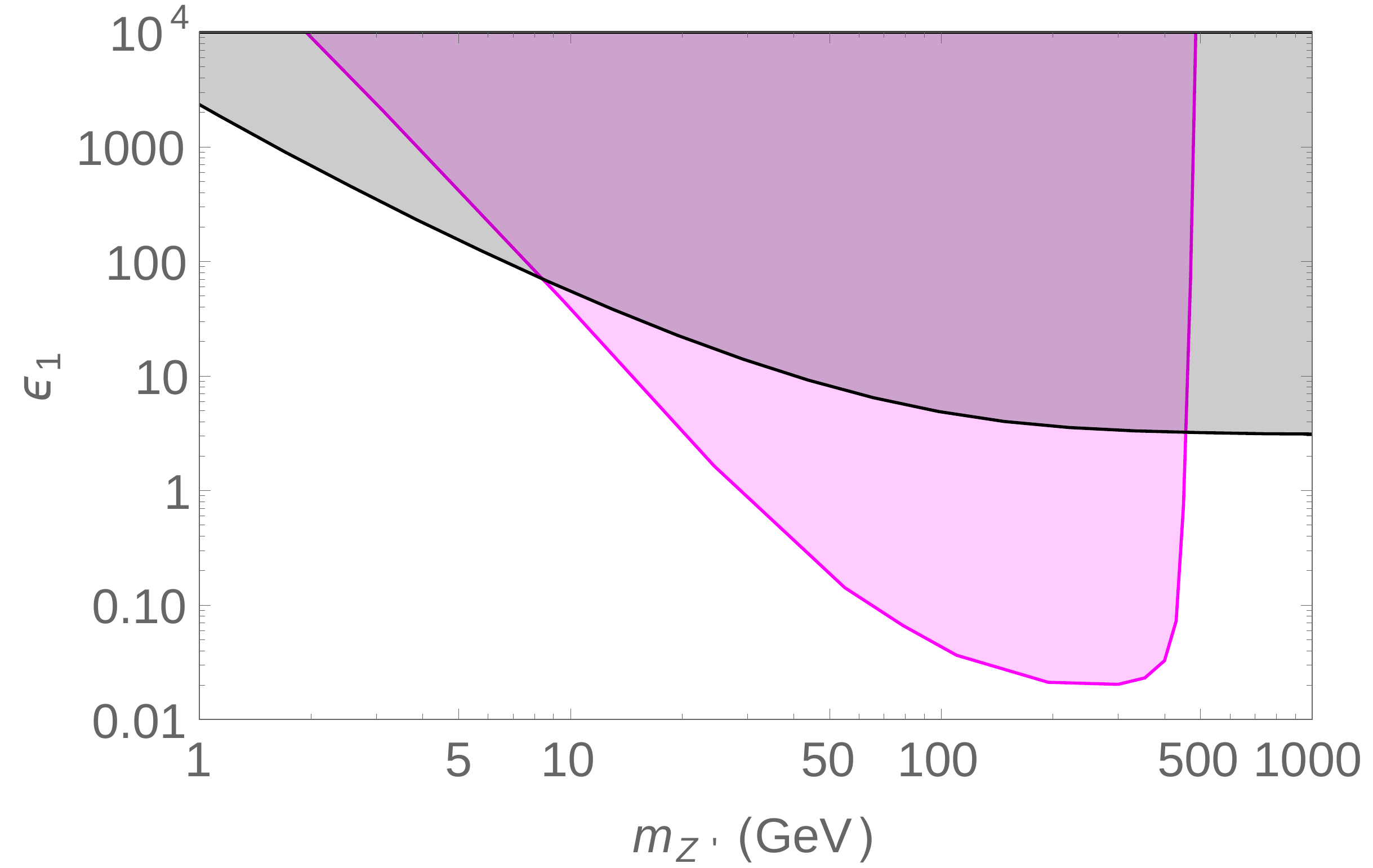}}
 \subfigure[]{
 \includegraphics[width=2.6in,height=2.2in, angle=0]{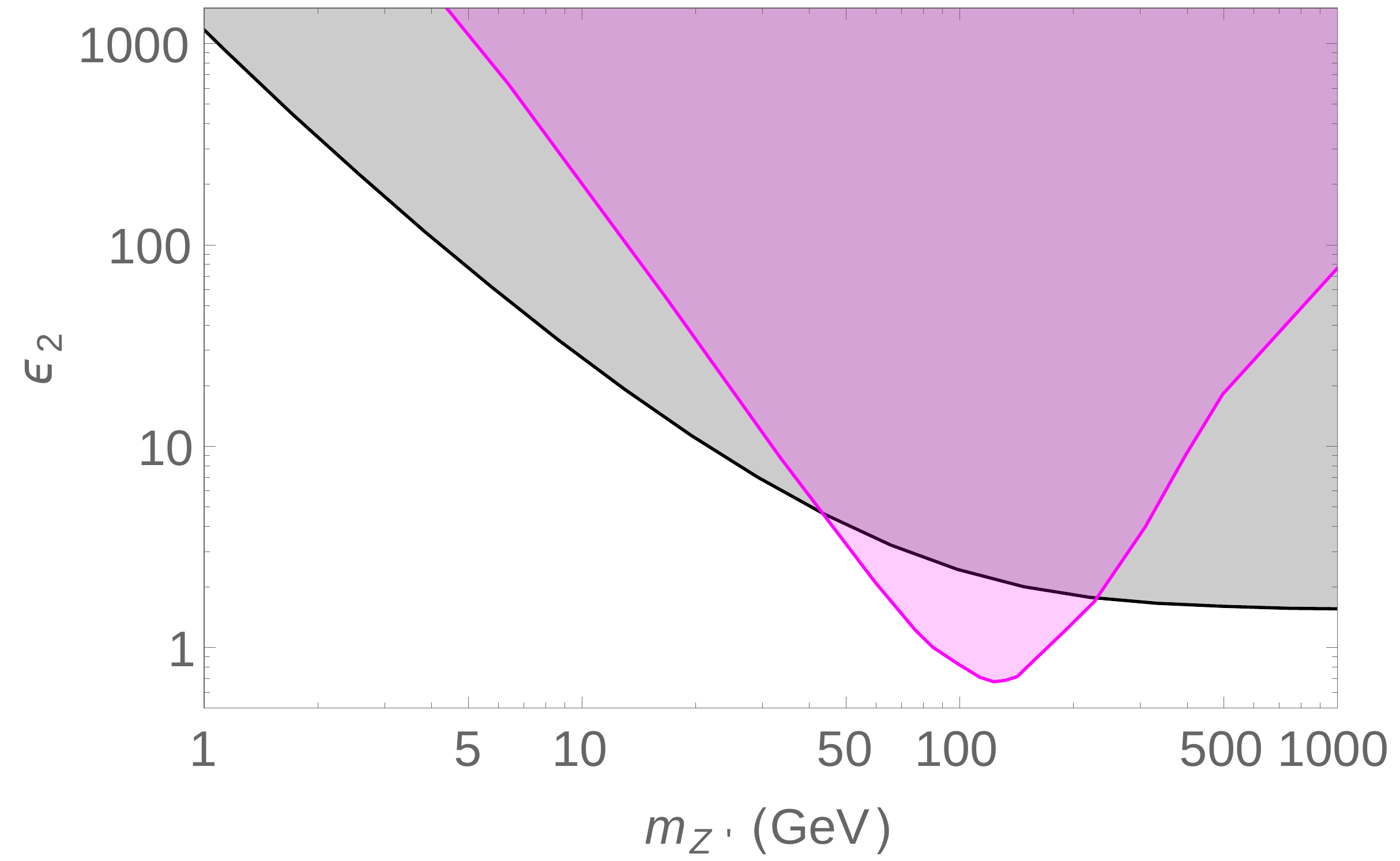}}   
 \caption{Constraints on $Z'$ induced NSI in presence of interactions expressed in eq.~\eqref{op1}, \eqref{op} and \eqref{opzp} respectively  as functions of $Z'$ mass. The pink and grey regions are excluded from LHC and IceCube respectively.}
 \label{fig:MZP}
\end{center}
 \end{figure}

\item 
Now, we consider a dim-5 interaction of neutrinos with $Z'$ with a dipole-like vertex structure,  
\bea
\mathcal{L} \supset  \frac{c^{(1)}}{\Lambda}(\bar{\nu_{i}^{c}} \sigma_{\mu \nu} P_L \n_{j})Z'^{\mu \nu} +g_{q} (\bar{q} \gamma^{\mu} q) Z'_{\mu} ,
\label{op}
\eea
where $\Lambda$ is the effective interaction scale. By demanding hermiticity of the Lagrangian, it can be noted that the term $(\bar{\nu_{i}^{c}} \sigma_{\mu \nu} P_L \n_{j})Z'^{\mu \nu}$ is non-vanishing only if $i \neq j$. Also, as shown for a renormalisable $Z'$ interaction in eq.~\eqref{op1}, the above interaction  also leads to kinetic mixing of $Z'$ with photon via a quark loop. This leads to transitional neutrino dipole moment $\mu_{ij}^M = (c^{(1)}/\Lambda) \epsilon_{loop} (k^2/(k^2-m_{Z'}^2)) $, where $k$ is the momentum of the photon and $\epsilon_{loop}$ is a loop factor expressed in eq.~\eqref{loopfac}.
The most stringent constraint on neutrino dipole moment comes from the  study of neutrino-electron scattering at Borexino and is given by, $\mu_{ij}^M \lsim 10^{-11}\mu_{B}$~\cite{Borexino:2017fbd}. For $m_{Z'} \sim$~MeV this leads to a rather stringent bound, $c^{(1)}g_{q}/\Lambda \lsim 10^{-5}$~GeV$^{-1}$. But   for a much heavier $Z'$ the constraint from Borexino becomes irrelevant: For $m_{Z'} = 5$~GeV, the Borexino bound  turns out to be  $c^{(1)} g_{q}/\Lambda \lsim 4.3 \times 10^3$~GeV$^{-1}$.

For $m_{Z'}=5$~GeV, LHC constraint from  monojet+$\slashed{E_{T}}$ search  turns out to be $c^{(1)} g_{q}/\Lambda \lsim 1.3 \times 10^{-3}$~GeV$^{-1}$,\ie $\epsilon_{1} \equiv (c^{(1)} g_{q} v/\Lambda)(v^2/2 m_{Z'}^2) \lsim 382$. 
IceCube constraint on this interaction, as can be followed from fig.~\ref{fig:dipole}, reads $c^{(1)} g_{q}/\Lambda \lsim 4.8 \times 10^{-4}$~GeV$^{-1}$,\ie $\epsilon_1 \lsim 143$, which is somewhat stronger than the constraints imposed by LHC.
For this interaction, the $m_{Z'}$ dependence of LHC and IceCube constraints on $\epsilon_1$ has been shown in fig.~\ref{fig:MZP}~(b). It can be seen that, the LHC constraint prevails the IceCube bound when $m_{Z'} \gtrsim 15$~GeV. However, due to the additional momentum enhancement, the width of $Z'$ becomes quite large in this case. Subsequently,  for $m_{Z'} \gtrsim 500$~GeV, the cross-section of $p p \rightarrow \nu \bar{\nu} j$ with this interaction does not significantly change with increasing couplings. This implies that there is no relevant constraint on this interaction for $m_{Z'} \gtrsim 500$~GeV from LHC.

\begin{figure}[h!]
 \begin{center}
 \includegraphics[width=3.7in,height=2.5in, angle=0]{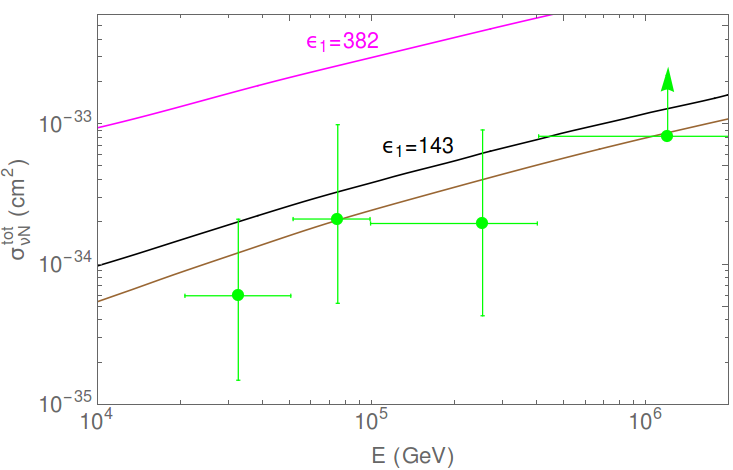}
 \caption{
Constraints on NSI described by  eq.~\eqref{op} in presence of a $Z'$ of mass 5~GeV. Colour coding is the same as in fig.~\ref{fig:simpleZp}. 
}
 \label{fig:dipole}
 \end{center}
 \end{figure}

\item 
Another dim-5 vertex for neutrino-$Z'$ interaction leading to neutrino-nucleon scattering can be written as, 
\bea
\mathcal{L} \supset  \frac{c^{(2)}}{\Lambda}(\bar{\nu^{c}} i \overset\leftrightarrow{\partial^{\m}} \nu )  Z'_{\mu}  + g_{q}(\bar{q} \gamma^{\mu} q) Z'_{\mu} ,
\label{opzp}
\eea
where $\Lambda$ is the effective interaction scale. 
As shown for the previous cases, the $\nu-$e scattering amplitude is suppressed by a loop factor $\epsilon_{loop}$, which renders the Borexino bound weaker than cases with tree-level electron-$Z'$ coupling. Thus, for the interaction in eq.~\eqref{opzp}, Borexino bound can be projected as,  $c^{(2)} g_{q}/\Lambda \lsim 3.4 \times 10^{4}$~GeV$^{-1}$ for $m_{Z'} = 5$~GeV.

Monojet+$\slashed{E_{T}}$ search at $\sqrt{s} = 13$~TeV at LHC leads to $c^{(2)} g_{q}/\Lambda \lsim 3.3 \times 10^{-3}$~GeV$^{-1}$,\ie $\epsilon_2 \equiv (c^{(2)} g_{q} v/\Lambda)(v^2/2 m_{Z'}^2) \lsim 982$  for $m_{Z'} = 5$ GeV, whereas the measurement of $\sigma_{\n N}^{tot}$ at IceCube provides a stronger bound, $c^{(2)} g_{q}/\Lambda \lsim 2.5 \times 10^{-4}$~GeV$^{-1}$,\ie $\epsilon_2 \lsim 75.6$.
The comparison of $\sigma_{\n N}^{tot}$ allowed from LHC and IceCube in the presence of the interaction given in eq.~\eqref{opzp}, is shown in fig.~\ref{fig:difZp}. The LHC and IceCube bounds for different values of $m_{Z'}$ have been depicted in fig.~\ref{fig:MZP}~(c) which shows that, the LHC bound becomes more significant than IceCube in the range $m_{Z'} \sim 40 - 220$~GeV.

\begin{figure}[h!]
 \begin{center}
 \includegraphics[width=3.7in,height=2.5in, angle=0]{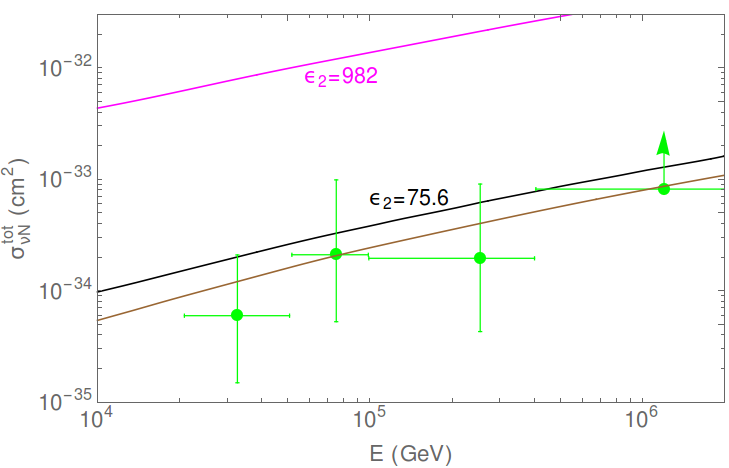}
 \caption{
Constraints on NSI described by eq.~\eqref{opzp} in presence of a $Z'$ of mass $5$~GeV. Colour coding is the same as in fig.~\ref{fig:simpleZp}.
 }
 \label{fig:difZp}
 \end{center}
 \end{figure}

\subsection{Contact type interactions}
\label{conNSI}

Neutrino-nucleon interaction can be realised via effective vertices which  lead to neutrino scattering off partons. In addition to the neutrino-quark operators, here we have also considered the case of neutrino-gluon effective interaction. In the following, we study the constraints on these effective interactions up to dim-7 from LHC and IceCube:

\item 
The dim-6 contact interaction leading to neutrino-quark scattering, which resembles the structure of the four-fermionic operator in eq.~\eqref{gen}, can be written as, 
\bea
\mathcal{L} \supset  \frac{c}{\Lambda^{2}}(\bar{\nu} \gamma_{\mu} \nu)  (\bar{q} \gamma^{\mu} q).
\label{opcon}
\eea

Here we use the notation, $\epsilon \equiv c v^2 /\Lambda^{2}$.
As mentioned in Sec.~\ref{exist}, a conservative constraint on the maximum allowed value of $\epsilon$ from low-energy neutrino DIS experiment CHARM is found to be, $\epsilon \sim 0.06$. Though, for different neutrino flavours, $\epsilon$ can take even higher values. 
We find that the LHC monojet+$\slashed{E_{T}}$ search leads to a somewhat stringent constraint, $\epsilon \lsim 0.02$ which is at par with the findings of refs.~\cite{Choudhury:2018azm,Franzosi:2015wha}.
The IceCube constraint from observation of cascade events is given as $-0.004 \lsim \epsilon \lsim 0.08$ and is shown in fig.~\ref{fig:con}. It is worth mentioning that, in presence of this interaction, the interference effect of the NSI and SM contributions in the process $pp \rightarrow \nu \bar{\n} j$ at LHC is rather small. This effect has been discussed in Appendix~\ref{Appendix:B} in more detail.

 This effective interaction can be interpreted as a dim-6 operator arising from an underlying renormalisable model consisting of a heavy $Z'$ with coupling to neutrinos and quarks at the tree-level as in eq.~\eqref{op1}. One can also formulate a tree-level matching condition, $\epsilon = (2 \sqrt{2} G_F)^{-1} (g_{Z'}/m_{Z'})^2$. But, in order to realise the maximum value of $\epsilon$ allowed by LHC in the underlying $Z'$ model, it would require a large coupling, $g_{Z'} \gtrsim 6$ with $m_{Z'} \gtrsim 8$~TeV~\cite{Choudhury:2018azm}. For such large values of $g_{Z'}$, the decay width of $Z'$ becomes larger than its mass and the aforementioned matching condition does not hold. Thus it is not sensible to match, or compare the bounds on this dim-6 interaction with the heavy $Z'$ model.


\begin{figure}[h!]
 \begin{center}
 \includegraphics[width=3.7in,height=2.5in, angle=0]{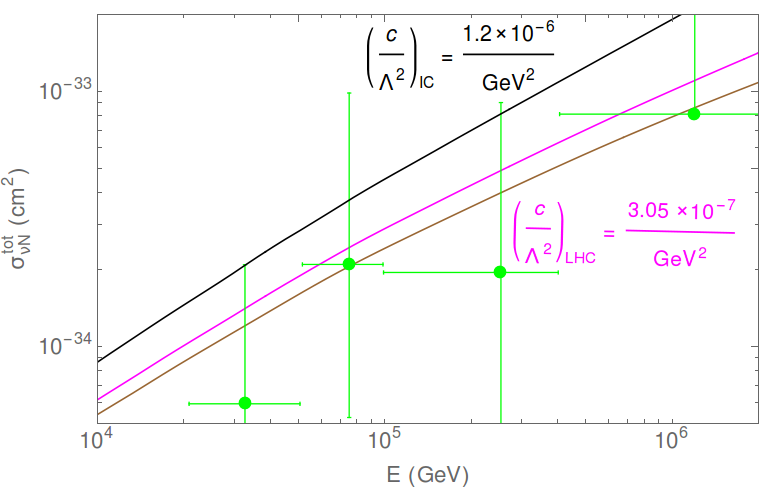}
 \caption{
 The maximum allowed values of NC neutrino-nucleon cross-section in presence of NSI appearing in eq.~\eqref{opcon}. Colour coding is the same as in fig.~\ref{fig:simpleZp}.}
 \label{fig:con}
 \end{center}
 \end{figure}

\item
\label{sub2} A dim-7 effective interaction which leads to neutrino-quark scattering is given by:
\bea
\mathcal{L} \supset \frac{c^{(3)}}{\Lambda^{3}}  \partial^{\nu}{(\bar{\nu_i^{c}} \sigma_{\mu \nu} P_L \n_j)} (\bar{q} \gamma^{\mu} q).
\label{op3} 
\eea
Here, in the same rationale as in eq.~\eqref{op},  $i\neq j$. Among the low energy experiments, the most stringent constraint on this interaction is imposed by CHARM, $c^{(3)}/\Lambda^3 \lsim 2.9 \times 10^{-7}$~GeV$^{-3}$~\cite{Altmannshofer:2018xyo}.
The LHC constraint on this interaction is found to be $c^{(3)}/\Lambda^3 \lsim 1.8 \times 10^{-10}$~GeV$^{-3}$.
Measurement of neutrino-nucleon scattering cross-section with IceCube cascade events gives a constraint, $c^{(3)}/\Lambda^3 \lsim 5.3 \times 10^{-8}$~GeV$^{-3}$.
Hence, the LHC bound is stronger than the CHARM and IceCube constraints. A comparison of the LHC and IceCube bounds can be followed from  fig.~\ref{fig:del}.

\begin{figure}[h!]
 \begin{center}
 \includegraphics[width=3.7in,height=2.5in, angle=0]{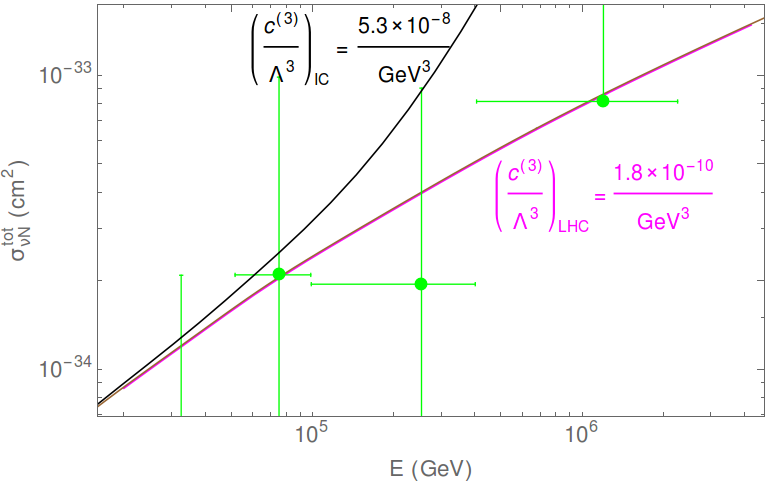}
 \caption{
 Constraints on NSI described by eq.~\eqref{op3}. Colour coding is the same as in fig.~\ref{fig:simpleZp}.}
 \label{fig:del}
 \end{center}
 \end{figure}

\item
\label{sub1} Another dim-7 effective Lagrangian for the neutrino-quark four-point interaction  is given by:
\bea
\mathcal{L} \supset \frac{c^{(4)}}{\Lambda^{3}} (\bar{\nu^{c}} i \overset\leftrightarrow{\partial_{\m}} \nu )(\bar{q} \gamma^{\mu} q).
\label{op31} 
\eea

The most relevant constraint among low energy experiments on this interaction comes from neutrino-nucleon scattering cross-section measurement at CHARM, $c^{(4)}/\Lambda^3 \lsim 1.2 \times 10^{-7}$~GeV$^{-3}$~\cite{Altmannshofer:2018xyo}.
Monojet+$\slashed{E}_T$ searches at LHC lead to a stronger constraint, $c^{(4)}/\Lambda^3 \lsim 8.6 \times 10^{-10}$~GeV$^{-3}$, whereas the bound from IceCube reads, $c^{(4)}/\Lambda^3 \lsim 2.6 \times 10^{-8}$~GeV$^{-3}$. 
As the last two cases, LHC provides a stronger constraint on this interaction compared to low-energy experiments and IceCube.
The neutrino-nucleon cross-sections at IceCube due to this interaction, corresponding to the upper limits of the IceCube and LHC constraints, are shown in fig.~\ref{fig:dif}.

\begin{figure}[h!]
 \begin{center}
 \includegraphics[width=3.7in,height=2.5in, angle=0]{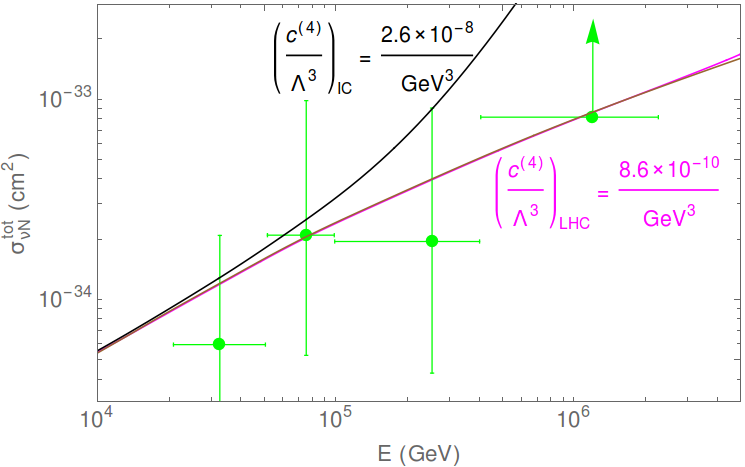}
 \caption{
Constraints on NSI appearing in eq.~\eqref{op31}. Colour coding is the same as  in fig.~\ref{fig:simpleZp}. }
 \label{fig:dif}
 \end{center}
 \end{figure}

\item
As mentioned earlier, neutrino-nucleon scattering can take place in the presence of effective interaction involving  neutrinos and gluons as well. A dim-7 term for such neutrino-gluon interaction is given as:
 \bea
\mathcal{L} \supset  \frac{c^{(5)}}{\Lambda^{3}} (\bar{\nu^{c}} P_L \n) G_{\mu \nu} G^{\mu \nu}.
\label{op2}
\eea

For the above interaction, the most relevant low-energy constraint comes from  the measurement of neutrino-nucleon cross-section  at CHARM,  $c^{(5)}/\Lambda^3 \lsim 1.6 \times 10^{-6}$~GeV$^{-3}$~\cite{Altmannshofer:2018xyo}.
LHC monojet+$\slashed{E}_T$ searches lead to the constraint, $c^{(5)}/\Lambda^3 \lsim 1.6 \times 10^{-10}$~GeV$^{-3}$. The neutrino-nucleon NC cross-section in presence of this interaction, with $\epsilon$ fixed at the upper bound obtained from LHC, is shown in fig.~\ref{fig:gluon}. The IceCube bound from the observation of cascade events is given by, $c^{(5)}/\Lambda^3 \lsim 5.5 \times 10^{-8}$~GeV$^{-3}$. 
Thus for the interaction given in eq.~\eqref{op2}, LHC gives a much stronger bound than both IceCube and CHARM.

A possible UV-completion of the operator in eq.~\eqref{op2} can be realised in the Type-II seesaw model, where an $SU(2)_L$ triplet~($\Delta$) with hypercharge-2 provides mass to the light neutrinos after it acquires a non-zero vacuum expectation value~(vev).
The measurement of the $T$-parameter renders $v_{\Delta}$ to be rather small, $v_{\Delta} < 4$~GeV. The lightest CP-even neutral component of the triplet, namely $\Delta^0$, mixes with the SM Higgs. The mixing parameter depends on the quartic couplings involving $H$ and $\Delta$, and the vev of the triplet as well. 
As the SM Higgs, $h$  has an effective coupling to a gluon pair through quark loops,  $h -\Delta^0$ mixing leads to an effective coupling of $\Delta^0$ to gluons too. 
 Thus the coefficient $c^{(5)}$ in eq.~\eqref{op2} is proportional to $y_{\nu} \sin {\a}$, where $y_{\nu}$ is the Yukawa coupling of neutrinos to $\Delta$, and $\alpha$ represents the mixing angle of $\Delta^0$ and the SM Higgs. The theoretical constraints, such as unitarity, stability, the measurement of $T$-parameter and $h \rightarrow \gamma \gamma$ constrain the value of $\sin \a$ significantly. The interplay of these bounds ensures that, for $m_H > 200$~GeV, $\sin \a \lesssim 0.02$~\cite{Das:2016bir}. Also, the Yukawa coupling leads to neutrino mass, $m_{\n} \sim y_{\n} v_{\Delta}$. Considering $m_{\n} \lesssim 0.1$~eV, $y_{\nu} \lesssim 10^{-10}$ for $v_{\Delta} = 1$~GeV and  $y_{\nu} \lesssim 10^{-6}$ for $v_{\Delta} = 10^{-4}$~GeV.
Thus the coefficient of this effective interaction is rather small if it is generated from such a renormalisable model and does not lead to a significant deviation from the SM value of NC neutrino-nucleon cross-section.

\begin{figure}[h!]
 \begin{center}
 \includegraphics[width=3.7in,height=2.5in, angle=0]{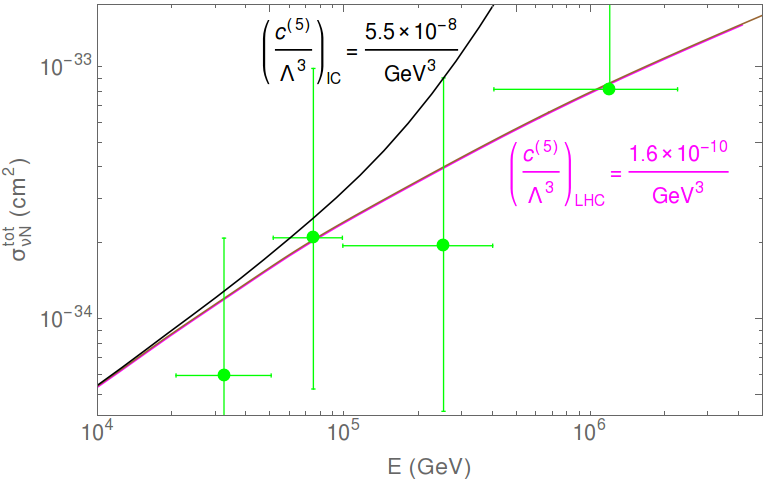}
\caption{
Constraints on NSI appearing in   eq.~\eqref{op2}. Colour coding is the same as in fig.~\ref{fig:simpleZp}.
}
 \label{fig:gluon}
 \end{center}
 \end{figure}

\end{enumerate}

We have found that, for the NSIs mediated by $Z'$ of  mass $m_{Z'} = 5$~GeV, IceCube provides a superior bound than LHC. 
Though, for contact-type NSI, the constraints from LHC are more significant than IceCube. 
However, due to the increasing nature of neutrino-nucleon cross-section in  the presence of contact-type NSI, as it can be seen from figs.~\ref{fig:con}, \ref{fig:del}, \ref{fig:dif} and \ref{fig:gluon}, the measurement of $\sigma_{\nu N}^{tot}$ in the bin $100-400$~TeV places the most stringent constraints on such interactions. 
An increase in the number of high energy neutrino events at IceCube-Gen2 will lead to reduced uncertainties in $\sigma_{\nu N}^{tot}$.   
For instance, the reduction in uncertainties in the bin $100 - 400$~TeV can improve the constraint on NSI appearing in eq.~\eqref{op31} by nearly a factor of two after 6 years of data from IceCube-Gen2.

No upgoing neutrinos have been observed in the energy range  $400 - 2004$~TeV. This leads to a lower bound on $\sigma_{\nu N}^{tot}$ which almost coincided with the SM prediction in this bin. 
Thus, any kind of new physics that leads to a substantial destructive interference with the SM contribution is  disfavoured from the energy bin $400 - 2004$~TeV. Note that, all the constraints from IceCube derived in this paper are independent of the sign of the couplings/Wilson coefficients as the NP contribution does not significantly interfere with SM, except the case described by eq.~\eqref{opcon}. For the case in eq.~\eqref{opcon}, only a small negative value is allowed from the IceCube due to the observation in the energy range  $400 - 2004$~TeV.
It is possible to distinguish the flavour of astrophysical neutrinos based on the CC interactions at the detector~\cite{Li:2016kra,Bustamante:2019sdb,Shoemaker:2015qul}, which in turn can lead to flavour-dependent constraints on NSI  of type $\bar{\nu} \nu \bar{q} q$. Thus, a better understanding of the neutrino flavour ratios at IceCube-Gen2 will also facilitate improved and flavour-dependent constraints on such NSI.

NSIs in eqs.~\eqref{op3}, \eqref{op31} and \eqref{op2} carry additional momentum dependence compared to the Fermi-type operator. In these cases, the neutrino-nucleon cross-sections increase with energy even faster, leading to more severe constraints from LHC. Moreover, as it can be seen from figs.~\ref{fig:dipole}-\ref{fig:gluon}, the value of $\sigma_{\nu N}^{tot}$ increases faster with neutrino energy in presence of  the contact-type interactions compared to the non-renormalisable interactions of a light $Z'$. This can be attributed to the propagator suppression in the $Z'$-mediated cases which relax the additional  momentum enhancement due to the non-renormalisable interactions. 

Low-energy experiments, such as MATHUSLA~\cite{Evans:2017lvd}, SHiP~\cite{Mermod:2017ceo}, FASER~\cite{Ariga:2019ufm}, dedicated to the search for new long-lived particles in the MeV-GeV range can put relevant constraints on the $Z'$ interactions considered in this paper. Such constraints, although flavour-dependent, can be stronger than that from IceCube, or even IceCube-Gen2. Though, these constraints only affect $Z'$ of mass $\lesssim 4$~GeV.

\section{Conclusion}
\label{concl}
NSIs lead to confusions in extracting the neutrino oscillation parameters by inflicting several degeneracies. Low-energy experiments provide constraints on the NSI parameters depending on the flavour structure. Among the high energy experiments, LHC leads to sizable constraints on the NSI parameters from generic new physics searches, in channels such as $p p \rightarrow j+\slashed{E}_T$. In IceCube, atmospheric neutrinos detected at DeepCore can also put flavour-specific constraints on NSI parameter at the level $\mathcal{O}(10^{-3})$. The observation of high energy astrophysical neutrinos at IceCube is particularly interesting in this context: It provides an opportunity to measure neutrino-nucleon cross-section at a value of $\sqrt{s}$ comparable to the LHC or even higher. This way it can also point to the existence of new physics at those high energies if in future, any deviation from the SM neutrino-nucleon cross-section is observed. The similarity in the centre-of-mass energies involved in concerned processes demands a comparative study of constraints on NSI from LHC and IceCube.

The uncertainty in neutrino flux can propagate in the  neutrino-nucleon cross-section extracted from the observation of astrophysical neutrinos.
High energy neutrinos reaching the IceCube from different directions traverse a different distance within the earth, providing sensitivity to the neutrino-nucleon cross-section which dictates the interaction length.   
Furthermore, the knowledge of parton distribution functions is also plagued with significant uncertainty for $E_{\nu} \gtrsim 10$~PeV. But the maximum energy for observed neutrinos goes up to $\sim$~PeV, for which the PDFs are well measured, primarily from HERA, thus making the error due to PDF irrelevant in light of current IceCube data. Thus, IceCube has enormous prospects for testing the non-standard neutrino interactions with high energy astrophysical neutrinos. In light of IceCube observations of shower and track events induced by such neutrinos,  estimates of neutrino-nucleon  scattering cross-section have been found in the literature~\cite{Aartsen:2017kpd, Bustamante:2017xuy}. As mentioned earlier, such a direct measurement of $\sigma_{\nu N}^{tot}$ can constrain the NSI parameters.

The non-standard interactions consisting of one charged lepton and one neutrino are constrained quite tightly from several low-energy experiments, EW precision tests, etc. We do not consider these kinds of interactions in our paper. 
Also, we are not interested in NSI involving two charged leptons and two partons, which suffer stringent constraints from various LEP measurements, meson decay etc. It has been mentioned that, it is possible to generate NSI of form $\bar{\nu}\nu \bar{q} q$ in a renormalisable model with a new vector boson $Z'$ without giving rise to the charged lepton counterpart of these interactions in the presence of new heavy fermions.
There also exist other scenarios where this can be attained, for example, in the presence of a specific gauge-invariant dim-8 operator. 
Though, if the NSIs are assumed to be generated from such operators with $d > 6$, the scale of new physics, $\Lambda$ can be lower than the case of dim-6 NSI. The implementation of IceCube bounds in this paper is based on an analysis which assumes equal neutrino flux across flavours. Thus, the constraints on NSI obtained in this paper are flavour-independent. 

We consider two subclasses of new interactions. Firstly, we discuss the case of a $Z'$ of mass $\sim \mathcal{O}(1)$~GeV with renormalisable and effective interactions up to dim-5. As mentioned earlier, in these cases, the IceCube bounds surpass the LHC constraints from  monojet+$\slashed{E}_T$ searches, which we illustrate for a $Z'$ with mass $m_{Z'} = 5$~GeV. 
In this context, future experiments dedicated to the search for new physics around $\sim 1$~GeV, such as MATHUSLA, SHiP, FASER, can put quite stringent constraints.  
The observation of coherent neutrino-nucleon scattering at COHERENT experiment can also lead to quite stringent, though flavour-specific, constraints in the presence of such a $Z'$~\cite{Abdullah:2018ykz,Liao:2017uzy,Kosmas:2017tsq}. We have also presented a comparison between LHC and IceCube bounds for different masses of $Z'$.
Broadly it has been seen that,  for $m_{Z'}$ within a few tens to a few hundreds of GeVs, the LHC  bounds are more significant than IceCube.  
 For example, with the renormalisable $Z'$ interactions as in eq.~\eqref{op1}, within the range $m_{Z'} \sim 35 - 500$~GeV, LHC provides  stronger constraints than IceCube. This also means, along with other new physics candidates like extra dimensions~\cite{Bustamante:2017xuy} and leptoquarks~\cite{Aartsen:2017kpd}, IceCube also has a remarkable discovery potential for $Z'$ of mass $\sim $~TeV.  Secondly, we take into account contact-type interactions involving two neutrinos and two partons up to dim-7. For such interactions, the LHC constraints are more significant than that from both IceCube and lower energy neutrino-scattering experiments.

The extraction of neutrino-nucleon cross-section is also affected by  astrophysical neutrino flux and flavour ratios. 
The constraints from IceCube derived in this paper can be improved in the upgraded version of this experiment, namely IceCube-Gen2, with a better understanding  of neutrino flux and flavour ratios~\cite{vanSanten:2017chb}. In case of discovery of even higher energy astrophysical neutrinos, the energy reach of IceCube can  supersede that of LHC. With current IceCube data, SM neutrino-nucleon cross-section is still allowed within $95\%$~CL. Any possible deviation from SM neutrino-nucleon cross-section may hint towards the existence of NP. 
With more statistics, it might also be possible to distinguish between different kinds of NSIs, if any, by studying the distribution of  high energy neutrino events  across deposited energy and zenith angle. 


\section{Acknowledgements}
S.~P. acknowledges email communication with Mauricio Bustamante regarding IceCube-Gen2 projected limits. This work is supported by the Department of Science and Technology, India {\it via} SERB grant EMR/2014/001177 and DST-DAAD grant INT/FRG/DAAD/P-22/2018.

\appendix

\section{Extracting constraints on NSI from CC neutrino-nucleon cross-section}
\label{implement}

 NC interactions of neutrinos of all flavours and CC interactions of $\nu_{\tau}$~($83\%$ times) and $\nu_e$ lead to cascade events at the IceCube detector.  Moreover, the interaction lengths of high energy neutrinos in earth depend upon the  neutrino-nucleon cross-section~(NC and CC). These make the extraction of $\sigma_{\nu N}^{tot}$ viable from the observation of cascade events at IceCube induced by high energy  neutrinos~\cite{Bustamante:2017xuy}. The NSIs considered in this paper provide additional contributions to the NC neutrino-nucleon cross-section, which can be constrained as, 
\bea
\sigma_{\nu N}^{NSI} \lesssim \sigma_{\nu N}^{tot, cas} -  \sigma_{\nu N}^{CC, IC} -  \sigma_{\nu N}^{NC, SM}.
\label{nsidef}
\eea
Here, $\sigma_{\nu N}^{tot, cas}$ denotes the total neutrino-nucleon cross-section measured from the IceCube observation of cascade events induced by high energy neutrinos~\cite{Bustamante:2017xuy}.
The second term in RHS of inequality~\eqref{nsidef},\ie the CC neutrino-nucleon cross-section,  $\sigma_{\nu N}^{CC, IC}$, can be measured rather precisely from the track events at IceCube~\cite{Aartsen:2017kpd}, so the related uncertainties are not implemented. This way one can estimate the remaining room for NSI contribution. 
Extracting the bound on $\sigma^{NSI}_{\nu N}$ in this way comes at the expense of introducing $\sim 2\%$ change in the neutrino flux compared to ref.~\cite{Bustamante:2017xuy}, which is even smaller than the effect of regeneration of high energy neutrinos passing through the earth. Considering the current uncertainties in  $\sigma_{\nu N}^{tot}$ found in ref.~\cite{Bustamante:2017xuy}, the effects of regeneration, which cause a change up to $\sim 10\%$ in the neutrino flux, does not have a significant impact on the estimated cross-section. By the same token, relevant bound on $\sigma^{NSI}_{\nu N}$ can be extracted using eq.~\eqref{nsidef}.

\vfill

\section{Differential cross-sections and interference effects}
\label{Appendix:B}

The differential cross-sections of the process $p p  \rightarrow \nu \bar{\nu} j$ for the contact NSI as in eq.~\eqref{opcon} can be written as the sum of contributions from the SM, NP and interference of these two: 
\bea
 \frac{d\sigma}{dp_{T} d\eta} = \frac{d\sigma_{SM}}{dp_{T} d\eta} + \frac{d\sigma_{int}}{dp_{T} d\eta} + \frac{d\sigma_{NP}}{dp_{T} d\eta},
\label{relativecontri}
\eea
 with,
\bea
\hspace{10pt}\frac{d\sigma_{SM}}{dp_{T} d\eta} &=& \frac{G_{F}^{2}}{\pi p_{T}} \Big( \frac{M_{Z}^4}{(Q_{tr}^{2}-M_{Z}^2)^2+ (\Gamma M_{Z})^2} \Big) Q_{tr}^2 \Big(1+ \frac{Q_{tr}^4}{(x_{1} x_{2} s)^{2}} \Big),\nn\\
\frac{d\sigma_{int}}{dp_{T} d\eta} &=& \frac{2 \epsilon G_{F}^{2}}{\pi p_{T}}  \Big( \frac{M_{Z}^2 (Q_{tr}^2-M_{Z}^2)}{(Q_{tr}^{2}-M_{Z}^2)^2+(\Gamma M_{Z})^2} \Big) Q_{tr}^2  \Big(1+ \frac{Q_{tr}^4}{(x_{1} x_{2} s)^{2}} \Big), \nn\\
\frac{d\sigma_{NP}}{dp_{T} d\eta} &=& \frac{\epsilon^{2} G_{F}^{2}}{\pi p_{T}}  Q_{tr}^2 \Big(1+ \frac{Q_{tr}^4}{(x_{1} x_{2} s)^{2}} \Big).
\label{B3}
\eea
Here, $x_1$ and $x_2$ are fractions of proton momentum transferred to the two initial partons involved in $p p  \rightarrow \nu \bar{\nu} j$ and $Q_{tr}$ is  momentum transferred to the neutrino pair.
 At LHC with  $\sqrt{s}=8$~TeV the cross-section for $pp \rightarrow \nu \bar{\n} j$ gets most of the contribution in  the $p_{T}$ range, $120 - 150$~GeV.
 To compare the relative contributions of the  different terms appearing in the RHS of eq.~\eqref{relativecontri}, we use the fact that, $\langle Q_{tr} \rangle \sim 500$~GeV for  $p_{T}=150$~GeV, $|\eta| < 2$ and $\sqrt{s}=8$~TeV~\cite{Busoni:2013lha}.
Using the second and third relations of eq.~\eqref{B3} one finds the ratio of the NP contribution to that from the interference term to be $\sim 2 M_{Z}^2/(\epsilon \langle Q_{tr}^2 \rangle)$. As $\langle Q_{tr} \rangle \sim 500$~GeV, this ratio turns out to be $\sim 0.33$ for the maximum allowed value of  $\epsilon \sim 0.19$. This implies, the interference term is subleading than the NP term in the cross-section of $p p \rightarrow \n \bar{\n} j$ with the dim-6 NSI term, which is somewhat opposite to the common perception. 
This happens due to an accidental conspiracy between  $\langle Q_{tr}^2 \rangle$ and current maximum allowed value of $\epsilon$. If the constraint on $\epsilon$ becomes even more stringent, with the value $\langle Q_{tr}^2 \rangle$ not changing significantly, the current picture can be reversed,\ie the interference term can be dominant over the NP contribution. A similar situation has been discussed in ref.~\cite{Friedland:2011za}.


\begin{thebibliography}{99}

\bibitem{Aartsen:2015knd}
  M.~G.~Aartsen {\it et al.} [IceCube Collaboration],
  Astrophys.\ J.\  {\bf 809} (2015) no.1,  98
  doi:10.1088/0004-637X/809/1/98
  [arXiv:1507.03991 [astro-ph.HE]].

\bibitem{TheIceCube:2016oqi} 
  M.~G.~Aartsen {\it et al.} [IceCube Collaboration],
  Phys.\ Rev.\ Lett.\  {\bf 117}, no. 7, 071801 (2016)
  doi:10.1103/PhysRevLett.117.071801
  [arXiv:1605.01990 [hep-ex]].
  
\bibitem{Coloma:2018idr} P.~Coloma, J.~Lopez-Pavon, I.~Martinez-Soler and H.~Nunokawa,
  Eur.\ Phys.\ J.\ C {\bf 78} (2018) no.8, 614
  doi:10.1140/epjc/s10052-018-6092-6
  [arXiv:1803.04438 [hep-ph]].
  
\bibitem{Hooper:2004xr} D.~Hooper, D.~Morgan and E.~Winstanley, 
  Phys.\ Lett.\ B {\bf 609} (2005) 206
  doi:10.1016/j.physletb.2005.01.034
  [hep-ph/0410094].

\bibitem{Esmaili:2018qzu}
  A.~Esmaili and H.~Nunokawa,
  Eur.\ Phys.\ J.\ C {\bf 79} (2019) no.1,  70
  doi:10.1140/epjc/s10052-019-6595-9
  [arXiv:1810.11940 [hep-ph]].

\bibitem{Liao:2016reh}
  J.~Liao and D.~Marfatia,
  Phys.\ Rev.\ Lett.\  {\bf 117} (2016) no.7,  071802
  doi:10.1103/PhysRevLett.117.071802
  [arXiv:1602.08766 [hep-ph]].

\bibitem{Liao:2018mbg}
  J.~Liao, D.~Marfatia and K.~Whisnant,
  Phys.\ Rev.\ D {\bf 99} (2019) no.1,  015016
  doi:10.1103/PhysRevD.99.015016
  [arXiv:1810.01000 [hep-ph]].
  
\bibitem{Moss:2017pur}
  Z.~Moss, M.~H.~Moulai, C.~A.~Argüelles and J.~M.~Conrad,
  Phys.\ Rev.\ D {\bf 97} (2018) no.5,  055017
  doi:10.1103/PhysRevD.97.055017
  [arXiv:1711.05921 [hep-ph]].

\bibitem{Denton:2018dqq}
  P.~B.~Denton, Y.~Farzan and I.~M.~Shoemaker,
  Phys.\ Rev.\ D {\bf 99} (2019) no.3,  035003
  doi:10.1103/PhysRevD.99.035003
  [arXiv:1811.01310 [hep-ph]].

\bibitem{Dey:2018yht}
  U.~K.~Dey, N.~Nath and S.~Sadhukhan,
  Phys.\ Rev.\ D {\bf 98} (2018) no.5,  055004
  doi:10.1103/PhysRevD.98.055004
  [arXiv:1804.05808 [hep-ph]].
  
\bibitem{Salvado:2016uqu}
  J.~Salvado, O.~Mena, S.~Palomares-Ruiz and N.~Rius,
  JHEP {\bf 1701} (2017) 141
  doi:10.1007/JHEP01(2017)141
  [arXiv:1609.03450 [hep-ph]].
  
\bibitem{Day:2016shw}
  M.~Day [IceCube Collaboration],
  J.\ Phys.\ Conf.\ Ser.\  {\bf 718} (2016) no.6,  062011.
  doi:10.1088/1742-6596/718/6/062011
  
\bibitem{Esmaili:2013fva}
  A.~Esmaili and A.~Y.~Smirnov,
  JHEP {\bf 1306} (2013) 026
  doi:10.1007/JHEP06(2013)026
  [arXiv:1304.1042 [hep-ph]].
  
\bibitem{Barranco:2010xt}
  J.~Barranco, O.~G.~Miranda, C.~A.~Moura, T.~I.~Rashba and F.~Rossi-Torres,
  JCAP {\bf 1110} (2011) 007
  doi:10.1088/1475-7516/2011/10/007
  [arXiv:1012.2476 [astro-ph.CO]].

\bibitem{Reynoso:2016hjr}
  M.~M.~Reynoso and O.~A.~Sampayo,
  Astropart.\ Phys.\  {\bf 82} (2016) 10
  doi:10.1016/j.astropartphys.2016.05.004
  [arXiv:1605.09671 [hep-ph]].

\bibitem{Arguelles:2017atb}
  C.~A.~Argüelles, A.~Kheirandish and A.~C.~Vincent,
  Phys.\ Rev.\ Lett.\  {\bf 119} (2017) no.20,  201801
  doi:10.1103/PhysRevLett.119.201801
  [arXiv:1703.00451 [hep-ph]].
  
\bibitem{deSalas:2016svi}
  P.~F.~de Salas, R.~A.~Lineros and M.~Tórtola,
  Phys.\ Rev.\ D {\bf 94} (2016) no.12,  123001
  doi:10.1103/PhysRevD.94.123001
  [arXiv:1601.05798 [astro-ph.HE]].
  
\bibitem{Huang:2018cwo}
  G.~Y.~Huang and N.~Nath,
  arXiv:1809.01111 [hep-ph].
  
\bibitem{Kelly:2018tyg}
  K.~J.~Kelly and P.~A.~N.~Machado,
  arXiv:1808.02889 [hep-ph].

\bibitem{Pandey:2018wvh}
  S.~Pandey, S.~Karmakar and S.~Rakshit,
  JHEP {\bf 1901} (2019) 095
  doi:10.1007/JHEP01(2019)095
  [arXiv:1810.04203 [hep-ph]].
  
\bibitem{Karmakar:2018fno}
  S.~Karmakar, S.~Pandey and S.~Rakshit,
  arXiv:1810.04192 [hep-ph].
  
\bibitem{Ng:2014pca}
  K.~C.~Y.~Ng and J.~F.~Beacom,
  Phys.\ Rev.\ D {\bf 90} (2014) no.6,  065035
   Erratum: [Phys.\ Rev.\ D {\bf 90} (2014) no.8,  089904]
  doi:10.1103/PhysRevD.90.065035, 10.1103/PhysRevD.90.089904

\bibitem{DiFranzo:2015qea}
  A.~DiFranzo and D.~Hooper,
  Phys.\ Rev.\ D {\bf 92} (2015) no.9,  095007
  doi:10.1103/PhysRevD.92.095007
  [arXiv:1507.03015 [hep-ph]].
 
\bibitem{Araki:2015mya}
  T.~Araki, F.~Kaneko, T.~Ota, J.~Sato and T.~Shimomura,
  Phys.\ Rev.\ D {\bf 93} (2016) no.1,  013014
  doi:10.1103/PhysRevD.93.013014
  [arXiv:1508.07471 [hep-ph]].

\bibitem{Mohanty:2018cmq}
  S.~Mohanty, A.~Narang and S.~Sadhukhan,
  JCAP {\bf 1903} (2019) no.03,  041
  doi:10.1088/1475-7516/2019/03/041
  [arXiv:1808.01272 [hep-ph]].

 \bibitem{Chauhan:2018dkd}
  B.~Chauhan and S.~Mohanty,
  Phys.\ Rev.\ D {\bf 98} (2018) no.8,  083021
  doi:10.1103/PhysRevD.98.083021
  [arXiv:1808.04774 [hep-ph]].

\bibitem{Shoemaker:2015qul}
  I.~M.~Shoemaker and K.~Murase,
  Phys.\ Rev.\ D {\bf 93} (2016) no.8,  085004
  doi:10.1103/PhysRevD.93.085004
  [arXiv:1512.07228 [astro-ph.HE]].

\bibitem{Cherry:2016jol}
  J.~F.~Cherry, A.~Friedland and I.~M.~Shoemaker,
  arXiv:1605.06506 [hep-ph]

\bibitem{Biggio:2009nt}
  C.~Biggio, M.~Blennow and E.~Fernandez-Martinez,
  JHEP {\bf 0908} (2009) 090
  doi:10.1088/1126-6708/2009/08/090
  [arXiv:0907.0097 [hep-ph]].

\bibitem{Bischer:2019ttk}
  I.~Bischer and W.~Rodejohann,
  arXiv:1905.08699 [hep-ph].

\bibitem{Carpentier:2010ue}
  M.~Carpentier and S.~Davidson,
  Eur.\ Phys.\ J.\ C {\bf 70} (2010) 1071
  doi:10.1140/epjc/s10052-010-1482-4
  [arXiv:1008.0280 [hep-ph]].
  
\bibitem{Aartsen:2017kpd}
  M.~G.~Aartsen {\it et al.} [IceCube Collaboration],
  Nature {\bf 551} (2017) 596
  doi:10.1038/nature24459
  [arXiv:1711.08119 [hep-ex]].
  
\bibitem{Falkowski:2017pss}
  A.~Falkowski, M.~González-Alonso and K.~Mimouni,
  JHEP {\bf 1708} (2017) 123
  doi:10.1007/JHEP08(2017)123
  [arXiv:1706.03783 [hep-ph]].

\bibitem{Bergmann:2000gp}
  S.~Bergmann, M.~M.~Guzzo, P.~C.~de Holanda, P.~I.~Krastev and H.~Nunokawa,
  Phys.\ Rev.\ D {\bf 62} (2000) 073001
  doi:10.1103/PhysRevD.62.073001
  [hep-ph/0004049].

\bibitem{Farzan:2017xzy}
  Y.~Farzan and M.~Tortola,
  Front.\ in Phys.\  {\bf 6} (2018) 10
  doi:10.3389/fphy.2018.00010
  [arXiv:1710.09360 [hep-ph]].
  
\bibitem{Davidson:2003ha}
  S.~Davidson, C.~Pena-Garay, N.~Rius and A.~Santamaria,
  JHEP {\bf 0303} (2003) 011
  doi:10.1088/1126-6708/2003/03/011
  [hep-ph/0302093].

\bibitem{Aaboud:2017buh}
  M.~Aaboud {\it et al.} [ATLAS Collaboration],
  JHEP {\bf 1710} (2017) 182
  doi:10.1007/JHEP10(2017)182
  [arXiv:1707.02424 [hep-ex]].

\bibitem{Gavela:2008ra}
  M.~B.~Gavela, D.~Hernandez, T.~Ota and W.~Winter,
  Phys.\ Rev.\ D {\bf 79} (2009) 013007
  doi:10.1103/PhysRevD.79.013007
  [arXiv:0809.3451 [hep-ph]].

\bibitem{Davidson:2011kr}
  S.~Davidson and V.~Sanz,
  Phys.\ Rev.\ D {\bf 84} (2011) 113011
  doi:10.1103/PhysRevD.84.113011
  [arXiv:1108.5320 [hep-ph]].

\bibitem{Aartsen:2014yll}
  M.~G.~Aartsen {\it et al.} [IceCube Collaboration],
  Phys.\ Rev.\ D {\bf 91} (2015) no.7,  072004
  doi:10.1103/PhysRevD.91.072004
  [arXiv:1410.7227 [hep-ex]].

\bibitem{Aartsen:2017xtt}
  M.~G.~Aartsen {\it et al.} [IceCube Collaboration],
  Phys.\ Rev.\ D {\bf 97} (2018) no.7,  072009
  doi:10.1103/PhysRevD.97.072009
  [arXiv:1709.07079 [hep-ex]].

\bibitem{Bustamante:2017xuy}
  M.~Bustamante and A.~Connolly,
  Phys.\ Rev.\ Lett.\  {\bf 122} (2019) no.4,  041101
  doi:10.1103/PhysRevLett.122.041101
  [arXiv:1711.11043 [astro-ph.HE]].

\bibitem{Dev:2019anc}
  P.~S.~Bhupal Dev {\it et al.},
  arXiv:1907.00991 [hep-ph].

\bibitem{Antusch:2008tz}
  S.~Antusch, J.~P.~Baumann and E.~Fernandez-Martinez,
  Nucl.\ Phys.\ B {\bf 810} (2009) 369
  doi:10.1016/j.nuclphysb.2008.11.018
  [arXiv:0807.1003 [hep-ph]].

\bibitem{Berezhiani:2001rs}
  Z.~Berezhiani and A.~Rossi,
  Phys.\ Lett.\ B {\bf 535} (2002) 207
  doi:10.1016/S0370-2693(02)01767-7
  [hep-ph/0111137].
  
\bibitem{Wolfenstein:1977ue}
  L.~Wolfenstein,
  Phys.\ Rev.\ D {\bf 17} (1978) 2369.
  doi:10.1103/PhysRevD.17.2369
  
\bibitem{Mikheev:1986gs}
  S.~P.~Mikheyev and A.~Y.~Smirnov,
  Sov.\ J.\ Nucl.\ Phys.\  {\bf 42} (1985) 913
   [Yad.\ Fiz.\  {\bf 42} (1985) 1441].

\bibitem{Esteban:2018ppq}
  I.~Esteban, M.~C.~Gonzalez-Garcia, M.~Maltoni, I.~Martinez-Soler and J.~Salvado,
  JHEP {\bf 1808} (2018) 180
  doi:10.1007/JHEP08(2018)180
  [arXiv:1805.04530 [hep-ph]].
  
\bibitem{Akimov:2017ade}
  D.~Akimov {\it et al.} [COHERENT Collaboration],
  Science {\bf 357} (2017) no.6356,  1123
  doi:10.1126/science.aao0990
  [arXiv:1708.01294 [nucl-ex]].

\bibitem{Borexino:2017fbd}
  M.~Agostini {\it et al.} [Borexino Collaboration],
  Phys.\ Rev.\ D {\bf 96} (2017) no.9,  091103
  doi:10.1103/PhysRevD.96.091103
  [arXiv:1707.09355 [hep-ex]].

\bibitem{CHARM}
  J.~Dorenbosch {\it et al.} [CHARM Collaboration],
  Phys.\ Lett.\ B {\bf 180} (1986) 303.
  doi:10.1016/0370-2693(86)90315-1
  
\bibitem{Denton:2018xmq}
  P.~B.~Denton, Y.~Farzan and I.~M.~Shoemaker,
  JHEP {\bf 1807} (2018) 037
  doi:10.1007/JHEP07(2018)037
  [arXiv:1804.03660 [hep-ph]].
  
\bibitem{Altmannshofer:2018xyo} 
  W.~Altmannshofer, M.~Tammaro and J.~Zupan,
  arXiv:1812.02778 [hep-ph].
  
\bibitem{Friedland:2011za}
  A.~Friedland, M.~L.~Graesser, I.~M.~Shoemaker and L.~Vecchi,
  Phys.\ Lett.\ B {\bf 714} (2012) 267
  doi:10.1016/j.physletb.2012.06.078
  [arXiv:1111.5331 [hep-ph]].
  
\bibitem{Choudhury:2018azm}
  D.~Choudhury, K.~Ghosh and S.~Niyogi,
  Phys.\ Lett.\ B {\bf 784} (2018) 248.
  doi:10.1016/j.physletb.2018.07.053

\bibitem{Alwall:2014hca}
  J.~Alwall {\it et al.},
  JHEP {\bf 1407} (2014) 079
  doi:10.1007/JHEP07(2014)079
  [arXiv:1405.0301 [hep-ph]].

\bibitem{Christensen:2008py}
  N.~D.~Christensen and C.~Duhr,
  Comput.\ Phys.\ Commun.\  {\bf 180} (2009) 1614
  doi:10.1016/j.cpc.2009.02.018
  [arXiv:0806.4194 [hep-ph]].

\bibitem{Sjostrand:2014zea}
  T.~Sj\"ostrand {\it et al.},
  Comput.\ Phys.\ Commun.\  {\bf 191} (2015) 159
  doi:10.1016/j.cpc.2015.01.024
  [arXiv:1410.3012 [hep-ph]].

\bibitem{Dercks:2016npn}
  D.~Dercks, N.~Desai, J.~S.~Kim, K.~Rolbiecki, J.~Tattersall and T.~Weber,
  Comput.\ Phys.\ Commun.\  {\bf 221} (2017) 383
  doi:10.1016/j.cpc.2017.08.021
  [arXiv:1611.09856 [hep-ph]].
  
\bibitem{ATLAS:2017dnw} The ATLAS collaboration [ATLAS Collaboration], 
  ATLAS-CONF-2017-060.

\bibitem{Aad:2015zva}
  G.~Aad {\it et al.} [ATLAS Collaboration],
  Eur.\ Phys.\ J.\ C {\bf 75} (2015) no.7,  299
   Erratum: [Eur.\ Phys.\ J.\ C {\bf 75} (2015) no.9,  408]
  doi:10.1140/epjc/s10052-015-3517-3, 10.1140/epjc/s10052-015-3639-7
  [arXiv:1502.01518 [hep-ex]].

\bibitem{Aartsen:2013jdh}
  M.~G.~Aartsen {\it et al.} [IceCube Collaboration],
  Science {\bf 342} (2013) 1242856
  doi:10.1126/science.1242856
  [arXiv:1311.5238 [astro-ph.HE]].
  
\bibitem{Aartsen:2013eka}
  M.~G.~Aartsen {\it et al.} [IceCube Collaboration],
  Phys.\ Rev.\ D {\bf 89} (2014) no.6,  062007
  doi:10.1103/PhysRevD.89.062007
  [arXiv:1311.7048 [astro-ph.HE]].

\bibitem{Gandhi:1998ri}
  R.~Gandhi, C.~Quigg, M.~H.~Reno and I.~Sarcevic,
  Phys.\ Rev.\ D {\bf 58} (1998) 093009
  doi:10.1103/PhysRevD.58.093009
  [hep-ph/9807264].

\bibitem{Aaron:2009aa}
  F.~D.~Aaron {\it et al.} [H1 and ZEUS Collaborations],
  JHEP {\bf 1001} (2010) 109
  doi:10.1007/JHEP01(2010)109
  [arXiv:0911.0884 [hep-ex]].

\bibitem{Abramowicz:2015mha}
  H.~Abramowicz {\it et al.} [H1 and ZEUS Collaborations],
  Eur.\ Phys.\ J.\ C {\bf 75} (2015) no.12,  580
  doi:10.1140/epjc/s10052-015-3710-4
  [arXiv:1506.06042 [hep-ex]].

\bibitem{Zenaiev:2015rfa}
  O.~Zenaiev {\it et al.} [PROSA Collaboration],
  Eur.\ Phys.\ J.\ C {\bf 75} (2015) no.8,  396
  doi:10.1140/epjc/s10052-015-3618-z
  [arXiv:1503.04581 [hep-ph]].
  
\bibitem{Gauld:2015yia}
  R.~Gauld, J.~Rojo, L.~Rottoli and J.~Talbert,
  JHEP {\bf 1511} (2015) 009
  doi:10.1007/JHEP11(2015)009
  [arXiv:1506.08025 [hep-ph]].
 
\bibitem{Cacciari:2015fta}
  M.~Cacciari, M.~L.~Mangano and P.~Nason,
  Eur.\ Phys.\ J.\ C {\bf 75} (2015) no.12,  610
  doi:10.1140/epjc/s10052-015-3814-x
  [arXiv:1507.06197 [hep-ph]].
  
\bibitem{Lai:2010vv}
  H.~L.~Lai, M.~Guzzi, J.~Huston, Z.~Li, P.~M.~Nadolsky, J.~Pumplin and C.-P.~Yuan,
  Phys.\ Rev.\ D {\bf 82} (2010) 074024
  doi:10.1103/PhysRevD.82.074024
  [arXiv:1007.2241 [hep-ph]].
  
\bibitem{Ohlsson:2012kf}
  T.~Ohlsson,
  Rept.\ Prog.\ Phys.\  {\bf 76} (2013) 044201
  doi:10.1088/0034-4885/76/4/044201
  [arXiv:1209.2710 [hep-ph]].
  
\bibitem{Farzan:2015doa}
  Y.~Farzan,
  Phys.\ Lett.\ B {\bf 748} (2015) 311
  doi:10.1016/j.physletb.2015.07.015
  [arXiv:1505.06906 [hep-ph]].
  
\bibitem{Miranda:2015dra}
  O.~G.~Miranda and H.~Nunokawa,
  New J.\ Phys.\  {\bf 17} (2015) no.9,  095002
  doi:10.1088/1367-2630/17/9/095002
  [arXiv:1505.06254 [hep-ph]].
    
\bibitem{Altmannshofer:2019zhy}
  W.~Altmannshofer, S.~Gori, J.~Martín-Albo, A.~Sousa and M.~Wallbank,
  arXiv:1902.06765 [hep-ph].
    
\bibitem{Huang:2017egl}
  G.~y.~Huang, T.~Ohlsson and S.~Zhou,
  Phys.\ Rev.\ D {\bf 97} (2018) no.7,  075009
  doi:10.1103/PhysRevD.97.075009
  [arXiv:1712.04792 [hep-ph]].
  
\bibitem{Dent:2012mx}
  J.~B.~Dent, F.~Ferrer and L.~M.~Krauss,
  arXiv:1201.2683 [astro-ph.CO].

\bibitem{Harnik:2012ni}
  R.~Harnik, J.~Kopp and P.~A.~N.~Machado,
  JCAP {\bf 1207} (2012) 026
  doi:10.1088/1475-7516/2012/07/026
  [arXiv:1202.6073 [hep-ph]].
    
\bibitem{Nelson:1989fx}
  A.~E.~Nelson and N.~Tetradis,
  Phys.\ Lett.\ B {\bf 221} (1989) 80.
  doi:10.1016/0370-2693(89)90196-2
  
\bibitem{Tulin:2014tya}
  S.~Tulin,
  Phys.\ Rev.\ D {\bf 89} (2014) no.11,  114008
  doi:10.1103/PhysRevD.89.114008
  [arXiv:1404.4370 [hep-ph]].
  
\bibitem{Aubert:2008as}
  B.~Aubert {\it et al.} [BaBar Collaboration],
  arXiv:0808.0017 [hep-ex].
    
\bibitem{Essig:2013vha}
  R.~Essig, J.~Mardon, M.~Papucci, T.~Volansky and Y.~M.~Zhong,
  JHEP {\bf 1311} (2013) 167
  doi:10.1007/JHEP11(2013)167
  [arXiv:1309.5084 [hep-ph]].

\bibitem{Achard:2001qw}
  P.~Achard {\it et al.} [L3 Collaboration],
  Phys.\ Lett.\ B {\bf 517} (2001) 75
  doi:10.1016/S0370-2693(01)01005-X
  [hep-ex/0107015].
  
\bibitem{Abdullah:2018ykz}
  M.~Abdullah, J.~B.~Dent, B.~Dutta, G.~L.~Kane, S.~Liao and L.~E.~Strigari,
  Phys.\ Rev.\ D {\bf 98} (2018) no.1,  015005
  doi:10.1103/PhysRevD.98.015005
  [arXiv:1803.01224 [hep-ph]].

\bibitem{Franzosi:2015wha}
  D.~Buarque Franzosi, M.~T.~Frandsen and I.~M.~Shoemaker,
  Phys.\ Rev.\ D {\bf 93} (2016) no.9,  095001
  doi:10.1103/PhysRevD.93.095001
  [arXiv:1507.07574 [hep-ph]].

\bibitem{Das:2016bir}
  D.~Das and A.~Santamaria,
  Phys.\ Rev.\ D {\bf 94} (2016) no.1,  015015
  doi:10.1103/PhysRevD.94.015015
  [arXiv:1604.08099 [hep-ph]].
  
\bibitem{Li:2016kra}
  S.~W.~Li, M.~Bustamante and J.~F.~Beacom,
  Phys.\ Rev.\ Lett.\  {\bf 122} (2019) no.15,  151101
  doi:10.1103/PhysRevLett.122.151101
  [arXiv:1606.06290 [astro-ph.HE]].
  
\bibitem{Bustamante:2019sdb}
  M.~Bustamante and M.~Ahlers,
  Phys.\ Rev.\ Lett.\  {\bf 122} (2019) 241101
  doi:10.1103/PhysRevLett.122.241101
  [arXiv:1901.10087 [astro-ph.HE]].
    
\bibitem{Evans:2017lvd}
  J.~A.~Evans,
  Phys.\ Rev.\ D {\bf 97} (2018) no.5,  055046
  doi:10.1103/PhysRevD.97.055046
  [arXiv:1708.08503 [hep-ph]].
  
\bibitem{Mermod:2017ceo}
  P.~Mermod [SHiP Collaboration],
  PoS NuFact {\bf 2017} (2017) 139
  doi:10.22323/1.295.0139
  [arXiv:1712.01768 [hep-ex]].
  
\bibitem{Ariga:2019ufm}
  A.~Ariga {\it et al.} [FASER Collaboration],
  arXiv:1901.04468 [hep-ex].
  
\bibitem{Liao:2017uzy}
  J.~Liao and D.~Marfatia,
  Phys.\ Lett.\ B {\bf 775} (2017) 54
  doi:10.1016/j.physletb.2017.10.046
  [arXiv:1708.04255 [hep-ph]].
  
\bibitem{Kosmas:2017tsq}
  D.~K.~Papoulias and T.~S.~Kosmas,
  Phys.\ Rev.\ D {\bf 97} (2018) no.3,  033003
  doi:10.1103/PhysRevD.97.033003
  [arXiv:1711.09773 [hep-ph]].
  
\bibitem{vanSanten:2017chb}
  J.~van Santen [IceCube Gen2 Collaboration],
  PoS ICRC {\bf 2017} (2018) 991.
  doi:10.22323/1.301.0991
  
\bibitem{Busoni:2013lha}
  G.~Busoni, A.~De Simone, E.~Morgante and A.~Riotto,
  Phys.\ Lett.\ B {\bf 728} (2014) 412
  doi:10.1016/j.physletb.2013.11.069
  [arXiv:1307.2253 [hep-ph]].
  
\end{thebibliography}
\end{document}